\pdfoutput=1
\documentclass[12pt]{article}
\catcode`\@=11
\@addtoreset{equation}{section}

\global\arraycolsep=2pt
\usepackage{mathrsfs,amsbsy,amssymb,latexsym,amsfonts,amsmath,cite,bm}
\usepackage{graphicx,color}
\usepackage{physics}
\usepackage{mathtools}
\usepackage[normalem]{ulem}
\usepackage{caption}
\usepackage{here}
\usepackage{subcaption}

\usepackage{enumerate}

\usepackage{hyperref}

\DeclareFontFamily{U}{BOONDOX-calo}{\skewchar\font=45 }
\DeclareFontShape{U}{BOONDOX-calo}{m}{n}{
  <-> s*[1.05] BOONDOX-r-calo}{}
\DeclareFontShape{U}{BOONDOX-calo}{b}{n}{
  <-> s*[1.05] BOONDOX-b-calo}{}
\DeclareMathAlphabet{\mathcalboondox}{U}{BOONDOX-calo}{m}{n}
\SetMathAlphabet{\mathcalboondox}{bold}{U}{BOONDOX-calo}{b}{n}
\DeclareMathAlphabet{\mathbcalboondox}{U}{BOONDOX-calo}{b}{n}

\allowdisplaybreaks

\begin{document}

\begin{flushright} 
RIKEN-iTHEMS-Report-26, KUNS-3116, STUPP-26-302
\end{flushright}
\vspace*{0.5cm}

\begin{center}
{\Large \bf Machine-Learning Search for Lax Connections
}
\vspace*{1.2cm} \\
{\large  Osamu Fukushima$^{\ast}$\footnote{E-mail:~osamu.fukushima$\_$at$\_$riken.jp}, 
Tomohiro Shigemura$^{\flat}$\footnote{E-mail:~shigemura$\_$at$\_$gauge.scphys.kyoto-u.ac.jp}, 
Ryosuke Suda$^{\ddagger}$\footnote{E-mail:~r.suda.813$\_$at$\_$ms.saitama-u.ac.jp}, \\
Norihiro Tanahashi$^{\flat}$\footnote{E-mail:~tanahashi$\_$at$\_$gauge.scphys.kyoto-u.ac.jp}
and Kentaroh Yoshida$^{\ddagger}$\footnote{E-mail:~kenyoshida$\_$at$\_$mail.saitama-u.ac.jp}
} 
\end{center}

\vspace*{0.2cm}

\begin{center}
$^{\ast}${\it iTHEMS, RIKEN, Wako, Saitama 351-0198, Japan}
\end{center}
\begin{center}
$^{\flat}${\it Department of Physics, Kyoto University, Kyoto 606-8502, Japan}
\end{center}
\begin{center}
$^{\ddagger}${\it Graduate School of Science and Engineering, Saitama University, \\ 
255 Shimo-Okubo, Sakura-ku, Saitama 338-8570, Japan}
\end{center}

\vspace{0.5cm}

\begin{abstract}
We apply a machine learning framework to search for Lax connections in two-dimensional non-linear sigma models using local current data. For the $SU(2)$ principal chiral model and the symmetric coset $S^2 = SU(2)/U(1)$, the method successfully recovers the full spectral-parameter families without using the known spectral curves as training targets. For the non-symmetric coset $T^{1,1}$, the optimization converges to reproducible low-loss maps that distill into a compact block-diagonal ansatz. However, analytic verification shows that this candidate is a ``fake Lax'' connection which satisfies on-shell flatness but fails to encode the two-dimensional equations of motion, whereas its point-particle reduction yields a genuine mechanical Lax pair. These results demonstrate that machine learning can effectively propose candidate ans\"atze and identify spectral structures, but low flatness loss alone does not certify genuine integrability, underscoring the necessity of analytic validation. 
\end{abstract}

\setcounter{footnote}{0}
\setcounter{page}{0}
\thispagestyle{empty}

\newpage

\tableofcontents

\renewcommand\thefootnote{\arabic{footnote}}

\section{Introduction}
\label{sec:introduction}

Classical integrability plays a fundamental role in the exact analysis of finite-dimensional dynamical systems and two-dimensional field theories. In field theory, this property is typically encoded in a Lax connection $L(\lambda)$, which depends on a continuous spectral parameter $\lambda$. Its flatness is equivalent to the classical equations of motion. The existence of a spectral-parameter family is essential: it allows the construction of monodromy matrices and an infinite tower of conserved quantities
\cite{abdalla2001non}. 
However, Lax representations are far from unique and usually rely on model-specific geometric structures or physical insights.
Constructing an appropriate Lax ansatz for less understood systems therefore remains a major challenge.

\medskip 

Recently, data-driven and optimization-based approaches have emerged as promising tools to search for Lax pairs and integrability computationally \cite{Krippendorf:2021lee,deKoster:2024akns,Adriazola:2025silo,Lin:2025laxpairfind,Adriazola:2026learning}.
These methods formulate Lax
compatibility, flatness, or related conservation-law constraints as numerical
objectives and search over parametrized operator or connection spaces.  At the
same time, compatibility alone can underdetermine the Lax representation and
may admit degenerate or anomalous candidates
\cite{Calogero:1991galore,Butler:2013fake,Sakovich:2020fake}.
This raises a central question for
field-theoretic searches: does a low-loss candidate belong to a genuine
spectral-parameter family whose flatness encodes the equations of motion, 
or is it merely an artifact of the optimization landscape?

\medskip 

Non-linear sigma models on group manifolds and coset spaces provide an ideal benchmark to address this question. Their field equations and geometric constraints are naturally expressed in terms of Lie-algebra-valued left invariant currents. While the principal chiral model (PCM) and symmetric coset sigma models like $S^2 = SU(2)/U(1)$ possess well-known Lax connections, non-symmetric cosets like $T^{1,1}$ generally lack such canonical constructions and often exhibit chaotic or non-integrable behavior\cite{Basu:2011di,Basu:2011fw,Asano:2015qwa,Kushiro:2022ksg}. They serve as a stringent testing ground for whether machine learning can propose meaningful ans\"atze without prior assumptions, and whether low flatness loss alone can serve as a diagnostic for classical integrability.  

\medskip 

In this paper, we apply a machine learning framework to search for Lax connections directly at the level of local current data in two-dimensional non-linear sigma models. First, for the $SU(2)$ principal chiral model and the symmetric coset $S^2$, we demonstrate that the optimization successfully recovers the standard continuous spectral-parameter families. 
The learned coefficient space maps out the exact algebraic curves defining the spectral dependence without manual intervention, validating the framework's capacity to detect one-parameter integrable structures.
We then extend the framework to the non-symmetric coset $T^{1,1}$ without assuming a prior Lax form. While the optimization robustly converges to reproducible low-loss maps that distill into a compact, block-diagonal ansatz, rigorous analytic verification reveals that this candidate is a fake Lax connection—it satisfies flatness identically due to algebraic identities, but its curvature condition fails to encode the two-dimensional equations of motion. Remarkably, reducing the system to geodesic motion on $T^{1,1}$ yields a genuine mechanical Lax pair, which we confirm both analytically and computationally.  

\medskip 

These findings highlight both the potential and the limitations of machine-learning-assisted discoveries. While optimization techniques provide an exceptionally powerful tool for exploring functional spaces, proposing candidate ans\"atze, and isolating spectral structures, a low flatness loss alone does not certify classical integrability. Rigorous analytic validation remains indispensable to distinguish true Lax representations from fake ones. 

\medskip 

This paper is organized as follows. Section~\ref{sec:pcm} describes the machine learning setup and the recovery of spectral-parameter families in the principal chiral model. Section~\ref{sec:symmetric-cosets} extends this analysis to symmetric coset models. Section~\ref{sec:t11-field} investigates the non-symmetric coset $T^{1,1}$, detailing the emergence of the block-diagonal map and the analytic proof of its fake Lax nature. Section~\ref{sec:t11-geodesic} presents the point-particle reduction and its mechanical Lax pair. Finally, Section~\ref{sec:summary} offers a summary and discussion.

\section{ML Search for Lax Connections in PCM}
\label{sec:pcm}

In this section, we study Lax connections in the principal chiral model (PCM) using a machine-learning framework. In particular, we elaborate on the method for determining the spectral parameter dependence of these Lax connections. 

\subsection{Classical integrability and Lax connections}

We begin with a brief review of the machine-learning framework introduced in \cite{Krippendorf:2021lee}, where the search for Lax connections is framed as an optimization problem. A key feature of this approach is its reliance on on-shell data sampled from the equations of motion, circumventing the need for full analytic solutions. Once an ansatz for the putative Lax structure is constructed, a loss function is minimized to systematically enforce the flatness condition.

\medskip 

In the following, we consider relativistic integrable field theories in $(1+1)$-dimensional Minkowski spacetime with coordinates $x^{\mu}=(x^0,x^1)=(t,x)$ and metric $\eta_{\mu\nu}=\mbox{diag}(-1,+1)$\,. 

\medskip 

Classical integrability is characterized by the existence of a Lax connection one-form
\begin{equation}
L(\lambda) = L_t(\lambda)\,dt + L_x(\lambda)\, dx\,, 
\end{equation}
such that it satisfies the flatness condition:
\begin{equation}
\partial_t L_x - \partial_x L_t + [L_t, L_x] = 0\,,
\label{flat}
\end{equation}
where $\lambda \in \mathbb{C}$ denotes the spectral parameter. 
For our purposes, it is convenient to work in light-cone coordinates, defined by
\begin{equation}
\sigma^\pm = t \pm x\,, \qquad \partial_\pm = \frac{1}{2}(\partial_t \pm \partial_x)\,.
\end{equation}
In these coordinates, the flatness condition \eqref{flat} is rewritten as
\begin{equation}
\partial_+ L_- - \partial_- L_+ + [L_+, L_-] = 0\,,
\end{equation}
where $L_\pm = (L_t \pm L_x)/2$\,. For later convenience, we also introduce the field strength two-form $F(\lambda)$ associated with the connection $L(\lambda)$, defined by
\begin{equation}F(\lambda) := dL(\lambda) + L(\lambda) \wedge L(\lambda)\,.
\end{equation}
The flatness condition can then be compactly expressed as $F(\lambda) = 0$\,.

\subsection{Lax connections in PCM}

As a prototypical example of relativistic integrable field theories, let us consider the principal chiral model (PCM). The PCM is a non-linear sigma model whose target space is defined by a Lie group $G$ itself.

\medskip 

It is useful to introduce the left-invariant (LI), Lie algebra
$\mathfrak{g}$-valued one-form defined by
\begin{equation}
J := g^{-1} dg\,.
\end{equation}
By construction, this current automatically satisfies the flatness condition:
\begin{equation}
\partial_+ J_- - \partial_- J_+ + [J_+, J_-] = 0\,. \label{eq:pcm-mc}
\end{equation}
In terms of the LI current, the classical action of the PCM is given by
\begin{equation}
S = -\frac{1}{2} \int \! d^2x\, \eta^{\mu\nu} \operatorname{Tr}(J_{\mu} J_{\nu})\,.
\end{equation}
Varying this action with respect to $g$ yields the equation of motion:
\begin{equation}
\partial_+ J_- + \partial_- J_+ = 0\,. \label{eq:pcm-eom}
\end{equation}
For later applications, it is useful to note that combining the equation of motion \eqref{eq:pcm-eom} and the flatness condition \eqref{eq:pcm-mc} leads to the following relations:
\begin{equation}
\partial_+ J_- = -\frac{1}{2} [J_+, J_-]\,, \qquad \partial_- J_+ = \frac{1}{2} [J_+, J_-]\,. \label{eq:pcm-eom-mc-combined}
\end{equation}

\medskip 

Finally, the one-parameter family of Lax connections associated with PCM is given by
\begin{equation}
L_+(\lambda) = \frac{1}{1-\lambda} J_+\,, \qquad L_-(\lambda) = \frac{1}{1+\lambda} J_-\,. 
\label{eq:pcm-standard-lax}
\end{equation}
This analytic expression provides a benchmark against which the output
of the machine-learning procedure adapted from
\cite{Krippendorf:2021lee} is evaluated.

\subsection{A map from LI current to Lax connection}

Before diving into the details of the machine-learning implementation, let us consider a linear map from the LI current $J_\pm$ to the Lax connection $L_\pm$, parameterized as
\begin{equation}
L_+ = a J_+\,, \qquad L_- = c J_-\,, \qquad a,~c \in \mathbb{C}\,. 
\label{eq:pcm-ac-ansatz}
\end{equation}
Here, $a$ and $c$ are arbitrary complex parameters to be determined by the machine-learning procedure. Crucially, these parameters are not independent due to the flatness condition. Indeed, a direct evaluation of the field strength yields
\begin{align}
F_{+-}(L) &= \partial_+ L_- - \partial_- L_+ + [L_+, L_-] \nonumber \\
&= c \partial_+ J_- - a \partial_- J_+ + ac [J_+, J_-] \nonumber \\
&= \left( -\frac{a}{2} - \frac{c}{2} + ac \right) [J_+, J_-]\,,
\end{align}
where we have utilized the relations \eqref{eq:pcm-eom-mc-combined} in the final equality. Therefore, the connection \eqref{eq:pcm-ac-ansatz} is on-shell flat if and only if
\begin{equation}
a + c - 2ac = 0\,. \label{eq:pcm-ac-curve}
\end{equation}
A standard parameterization of this constraint curve is given by
\begin{equation}
a = \frac{1}{1-\lambda}\,, \qquad c = \frac{1}{1+\lambda}\,. \label{eq:pcm-ac-lambda}
\end{equation}
A key observation is that, within the ansatz
\eqref{eq:pcm-ac-ansatz}, the standard PCM Lax family appears as the
one-dimensional flatness locus \eqref{eq:pcm-ac-curve}. A single
optimization run selects one point on this locus, while runs from
different initializations can collectively recover the curve through
the distribution of their learned endpoints. The resulting spread
should therefore not be interpreted as arbitrary numerical scatter:
its organization along \eqref{eq:pcm-ac-curve} is the expected
signature of the one-parameter family.

\subsection{Machine-learning setup}

We now formulate the machine-learning problem in the simplest PCM setting.
This example will serve as the prototype for the coset cases discussed in
the following sections.

\paragraph{Ansatz specification.}
In the general strategy of \cite{Krippendorf:2021lee}, the components of a
candidate Lax connection may be represented by neural networks depending on
local current data and, if necessary, their derivatives:
\begin{equation}
L_\pm
=
L_{\pm,\mathrm{NN}}
(J_+,J_-,\partial_+J_-,\partial_-J_+,\ldots;\theta),
\end{equation}
where $\theta$ denotes the trainable parameters.  For the PCM benchmark,
however, the analytic structure suggests the minimal current-level ansatz
\eqref{eq:pcm-ac-ansatz},
\begin{equation}
L_+(J_+;a)=aJ_+,\qquad
L_-(J_-;c)=cJ_-,
\end{equation}
with two scalar coefficients $a,c$ to be optimized.

\paragraph{On-shell data generation.}
We generate training data by sampling local values of the currents
$(J_+,J_-)$ in the Lie algebra.  The derivative data are then fixed by the
PCM equation of motion together with the Maurer--Cartan relation.  Using
\eqref{eq:pcm-eom-mc-combined}, we set
\begin{equation}
\partial_+J_-=-\frac12[J_+,J_-],
\qquad
\partial_-J_+=\frac12[J_+,J_-].
\end{equation}
Thus every sample is an on-shell local current configuration.

\paragraph{Loss function.}
The training objective is the on-shell flatness of the candidate connection.
For a batch of $N$ samples, we use
\begin{equation}
\mathcal L_{\rm flat}
=
\frac{1}{N}
\sum_{n=1}^{N}
\bigl\|
F_{+-}(L)
\bigr\|^2
=
\frac{1}{N}
\sum_{n=1}^{N}
\bigl\|
\partial_+L_-^{(n)}
-\partial_-L_+^{(n)}
+[L_+^{(n)},L_-^{(n)}]
\bigr\|^2 ,
\label{eq:pcm-loss}
\end{equation}
where the standard Frobenius norm for a matrix $A$ is defined as 
\begin{equation}
    \|A\|^2 = \sum_{i,j}|A_{ij}|^2\,.
    \label{eq:matrix-norm}
\end{equation}
The derivatives of $L_\pm$ are obtained by substituting the on-shell
derivatives of the currents, and the commutator is evaluated directly in the
matrix representation of the Lie algebra.

\medskip 

For the minimal ansatz above, the curvature reduces to
\begin{equation}
\partial_+L_--\partial_-L_+ + [L_+,L_-]
=
\left(-\frac{a}{2}-\frac{c}{2}+ac\right)[J_+,J_-].
\end{equation}
Therefore, up to an overall normalization independent of $a,c$,
\begin{equation}
\mathcal L_{\rm flat}
=
\bigl|a+c-2ac\bigr|^2
\frac{1}{N}
\sum_{n=1}^{N}
\bigl\|[J_+^{(n)},J_-^{(n)}]\bigr\|^2 .
\label{eq:pcm-loss-reduced}
\end{equation}
Thus, in this reduced PCM experiment, learning the Lax connection is
equivalent to finding the flatness locus
\eqref{eq:pcm-ac-curve}
rather than fitting a prescribed value of the spectral parameter.

\subsection{Real Slice Used for Visualization}

For simplicity, we consider the simplest case $G=SU(2)$\,. 

\medskip

In the actual PCM Lax connection, the coefficients are naturally allowed to be
complex. This is also how the setup should be viewed in principle.
For the purpose of visualization, however, it is useful to restrict the optimization
to the real slice
\begin{equation}
a,c\in\mathbb{R}.
\end{equation}
Under this restriction, the learned endpoints can be plotted in the $(a,c)$-plane,
where they are expected to lie on the real section of the spectral-parameter curve
\eqref{eq:pcm-ac-curve}.

\medskip

\begin{figure}[t]
\centering
\includegraphics[width=.72\textwidth]{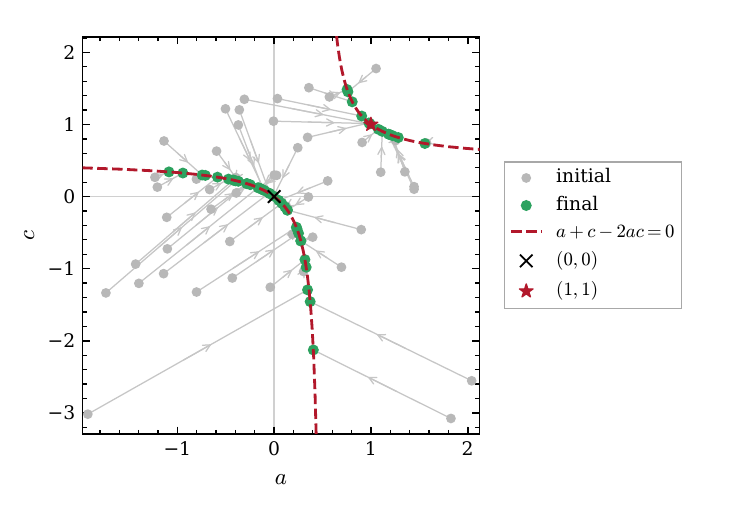}
\caption{Scatter plot of learned parameters $(a,c)$ for the PCM with the restriction
$a,c\in\mathbb{R}$. Gray points denote initial values and green points final values.
The dashed red curve is the exact on-shell flatness locus
$a+c-2ac=0$, namely the real slice of the spectral-parameter family.}
\label{fig:pcm-ac}
\end{figure}

Figure~\ref{fig:pcm-ac} shows a representative scatter plot obtained in this way.
Different random initializations lead to different final values, but these values
cluster on the algebraic curve
\eqref{eq:pcm-ac-curve}.
This gives a direct numerical confirmation that the optimization is detecting the
spectral-parameter family rather than selecting an arbitrary isolated solution.

\subsection{Towards Explicit Learning of the Spectral Parameter}
\label{sec:pcm-lambda-body}

While the previous discussion treats the spectral parameter indirectly by learning individual points on the family, a natural next step is to introduce $\lambda$ as an explicit functional input:
\begin{equation}
    L_\pm = L_\pm(J_+, J_-; \lambda).
\end{equation}
Training the model with explicit $\lambda$ labels is non-trivial. A naive implementation may collapse to a $\lambda$-independent solution, thereby losing the essential integrable structure. To prevent this, non-trivial $\lambda$-dependence must be enforced through architectural constraints, training set design, or additional loss terms.

\medskip

We now report a realization of this program for the PCM; the
setup and diagnostics are collected in
Appendix~\ref{app:pcm-lambda}. We keep the current-level ansatz
\eqref{eq:pcm-ac-ansatz} but promote its coefficients to \emph{functions} of the
spectral parameter, $a=a(\lambda)$ and $c=c(\lambda)$, represented by a small
neural network $\mathcal N_\theta$ taking
$(\operatorname{Re}\lambda,\operatorname{Im}\lambda)$ as input.%
\footnote{As explained below, $a(\lambda)$ is prescribed by hand, so that only $c(\lambda)$ is expressed by the network.}
The quantity minimized is the flatness residual itself: at each sampled $\lambda$ the
connection $L_\pm$ is assembled from the sampled on-shell currents and
$\|F_{+-}\|^2$ is evaluated
as in
\eqref{eq:pcm-loss}.
This residual is minimized as the sole data term,
averaged over randomly sampled points in 
a given domain in the complex $\lambda$-plane
and over on-shell currents.
If the training succeeds, the network should reproduce the spectral curve~\eqref{eq:pcm-ac-curve} in the above complex domain of $\lambda$.

\medskip

To accomplish the above program, we need to overcome the following two obstacles.
\begin{enumerate}
    \item 
        \textit{Reparametrization degree of freedom}.
        Flatness imposes the relation \eqref{eq:pcm-ac-curve} between $a$ and $c$ at each $\lambda$, but it does not fix their dependence on $\lambda$ uniquely.
        We fix this freedom by prescribing the functional form of one coefficient,
        which we choose to be $a(\lambda)$.
        The standard parametrization
        $a(\lambda)=1/(1-\lambda)$, shown in \eqref{eq:pcm-ac-lambda},
        presupposes knowledge of the spectral curve~\eqref{eq:pcm-ac-curve}.
        We instead make the more solution-agnostic choice
        $a(\lambda)=\lambda$; such an approach may be applicable to general problems whose solutions are unknown in advance.
        The choice $a(\lambda)=\lambda$ solves the curve
        \eqref{eq:pcm-ac-curve} for the definite analytic solution
        $c_{\rm an}(\lambda)=\lambda/(2\lambda-1)$.
        To avoid problems arising from the pole in $c(\lambda)$ at $\lambda=1/2$, we restrict the training domain to a rectangle
        $\operatorname{Re}\lambda\in[-2,0]$,
        $\operatorname{Im}\lambda\in[-1,1]$ in $\mathbb{C}$.

    \item
        \textit{Uneven training signal.} On shell every term of $F_{+-}$ is
        proportional to $[J_+,J_-]$, so the bare loss in its on-shell form
        \eqref{eq:pcm-loss-reduced} carries
        an overall factor $\|[J_+,J_-]\|^2$, which fluctuates strongly across the current
        batch. The gradient is then dominated by the largest-amplitude configurations
        rather than by the information in $a+c-2ac$. We equalize the training signal by using a 
        loss function normalized by the component norm given by
        \begin{equation}
          \mathcal L_{\rm cn}
          =\left\langle
             \frac{\|F_{+-}\|^2}
                  {\|\partial_+L_-\|^2+\|\partial_-L_+\|^2+\|[L_+,L_-]\|^2+\varepsilon}
           \right\rangle,
          \qquad \varepsilon\ll1\,.
          \label{eq:pcm-loss-cn}
        \end{equation}
        The denominator makes every current configuration contribute
        to the gradient with comparable strength regardless of its amplitude.
        Also, the normalization is problem-agnostic in the sense that it is built only from quantities already needed for $F_{+-}$,
        so it may be applied as is to other problems. Indeed, the same normalization is used in the coset examples discussed in section~\ref{sec:priorfree-body}.
\end{enumerate}

\begin{figure}[t]
\centering
\includegraphics[width=.5\textwidth]{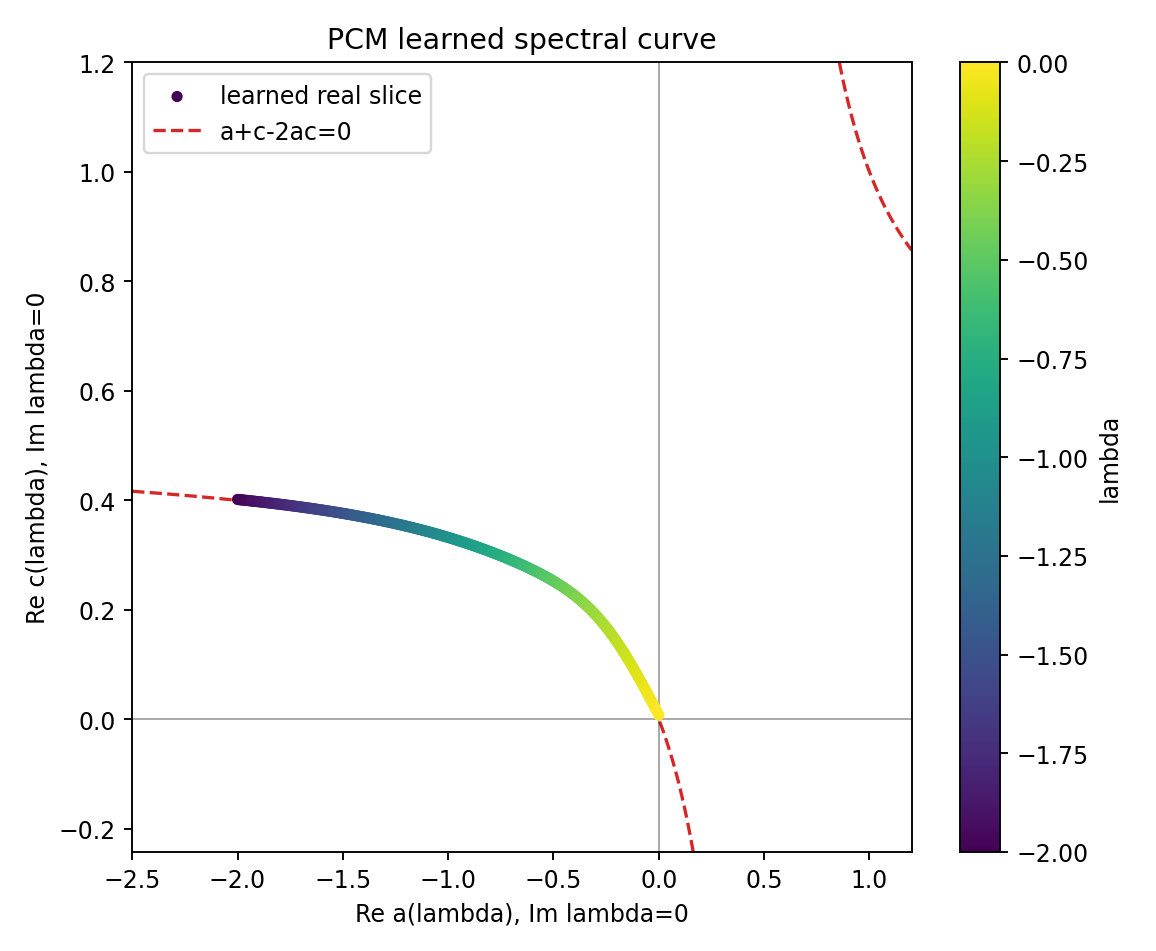}
\caption{Learned coefficients $(a(\lambda),c(\lambda))$ (real-axis slice, color
$=\operatorname{Re}\lambda$) overlaid on the analytic spectral curve
$a+c-2ac=0$ (red dashed), for the fixed-parametrization run ($a=\lambda$,
$10^4$ steps) trained with the component-norm loss function. The network reproduces
the curve as a genuine function of $\lambda$ across the whole sampled rectangle
$\operatorname{Re}\lambda\in[-2,0]$, $\operatorname{Im}\lambda\in[-1,1]$; the
deviation of $c$ from $\lambda/(2\lambda-1)$ over the complex domain has mean
modulus $3.0\times10^{-3}$ and maximum $2.4\times10^{-2}$
(see Appendix~\ref{app:pcm-lambda}).}
\label{fig:pcm-lambda-body}
\end{figure}

Figure~\ref{fig:pcm-lambda-body} (and \ref{fig:pcm-lambda-3d} in Appendix~\ref{app:pcm-lambda})
show that the network recovers the
spectral-parameter dependence itself, not merely isolated members of the
family: minimizing the flatness residual alone
with $a(\lambda)=\lambda$ supplied
yields $c(\lambda)$ that follows $\lambda/(2\lambda-1)$ to
a mean absolute error of $\sim\!3\times10^{-3}$ across the sampled rectangle.
The setup and the diagnostics behind
this number
are given in
Appendix~\ref{app:pcm-lambda}.

\section{Symmetric coset sigma models}
\label{sec:symmetric-cosets}

In this section, we extend the procedure developed for the PCM in \cite{Krippendorf:2021lee} to symmetric coset sigma models.

\subsection{Lax connections in symmetric coset sigma models}

It is known that a 2D non-linear sigma model whose target space is a symmetric coset, which is referred to as a symmetric coset sigma model, is classically integrable, as guaranteed by the existence of the associated Lax connection. For details of symmetric coset sigma models, see, for example,  \cite{Yoshida:2021qfl}. 

\medskip 

First, let us introduce the geometry of symmetric cosets. Consider a coset $M=G/H$\,, where $G$ is a Lie group and $H$ is its subgroup. 
Let $\mathfrak{g}$ and $\mathfrak{h}$ be the Lie algebras of $G$ and $H$, respectively. Then, $\mathfrak{g}$ can be decomposed as the vector space direct sum
\begin{equation}
\mathfrak g=\mathfrak h\oplus\mathfrak m\,,
\end{equation}
where $\mathfrak{m}$ is the orthogonal complement to $\mathfrak{h}$\,. If $\mathfrak{h}$ and $\mathfrak{m}$ satisfy the following $\mathbb{Z}_2$-grading relations:
\begin{equation}
[\mathfrak h,\mathfrak h]\subset\mathfrak h\,,
\qquad
[\mathfrak h,\mathfrak m]\subset\mathfrak m\,,
\qquad
[\mathfrak m,\mathfrak m]\subset\mathfrak h\,,
\label{eq:sym-z2}
\end{equation}
then the coset $M=G/H$ is called a symmetric coset. 
The grading is defined such that $\mathfrak{h}$ has grade 0 and $\mathfrak{m}$ has grade 1.

\medskip 

When we consider a symmetric coset, the LI one-form can be decomposed as 
\begin{equation}
J_\pm=g^{-1}\partial_\pm g=J_\pm^{(0)}+J_\pm^{(1)}\,,
\qquad
J_\pm^{(0)}\in\mathfrak h\,,
\quad
J_\pm^{(1)}\in\mathfrak m\,. \label{decomp}
\end{equation}
Here, $J_{\pm}^{(0)}$ corresponds to the gauge degrees of freedom. According to this grading property, the flatness condition of the LI one-form can be decomposed as follows:
\begin{eqnarray}
  && \partial_+ J_-^{(0)} - \partial_- J_+^{(0)} +[J_+^{(0)}, J_-^{(0)}] + [J_+^{(1)}, J_-^{(1)}]  = 0\,, \label{scsm-flat1}
  \\ 
 && \partial_+ J_-^{(1)} - \partial_- J_+^{(1)} 
 + [J_+^{(0)}, J_-^{(1)}] + [J_+^{(1)}, J_-^{(0)}] = 0\,.   \label{scsm-flat2}
\end{eqnarray}

\medskip 

The classical action of the symmetric sigma model is given by 
\begin{equation}
S = -\frac{1}{2} \int \! d^2x\, \eta^{\mu\nu} \operatorname{Tr}(J^{(1)}_{\mu} J^{(1)}_{\nu})\,.
\end{equation}
Taking a variation of this action, the equation of motion is obtained as
\begin{equation}
\partial_+ J_-^{(1)} + [J_+^{(0)},J_-^{(1)}] +\partial_- J_+^{(1)} + [J_{-}^{(0)},J_+^{(1)}] =0\,. 
\label{scsm-eom}
\end{equation}
Note that the equation of motion \eqref{scsm-eom} together with the flatness condition \eqref{scsm-flat2} are equivalent to:
\begin{eqnarray}
    && \partial_+ J_-^{(1)} + [J_+^{(0)},J_-^{(1)}] = 0\,, \label{rel1} \\ 
    && \partial_- J_+^{(1)} + [J_{-}^{(0)},J_+^{(1)}]=0\,,
    \label{rel2} 
\end{eqnarray}
which are useful for generating current data and evaluating the flatness condition of the Lax connection.

\medskip

According to the decomposition \eqref{decomp}, the standard one-parameter family of Lax connections is given by 
\begin{equation}
L_+(\lambda)=J_+^{(0)}+\lambda J_+^{(1)}\,,
\qquad
L_-(\lambda)=J_-^{(0)}+\lambda^{-1}J_-^{(1)}\,.
\label{eq:sym-standard-lax}
\end{equation}
The flatness of this Lax connection can be evaluated as 
\begin{align}
& \partial_+ L_- - \partial_- L_+ + [L_+, L_-] \nonumber \\ 
={}& \partial_+ J_-^{(0)} - \partial_- J_+^{(0)} + \lambda^{-1} \partial_+ J_-^{(1)} - \lambda\,\partial_- J_+^{(1)} + [J_+^{(0)}, J_-^{(0)}]  + [J_+^{(1)}, J_-^{(1)}] \nonumber \\ 
& + \lambda [J_{+}^{(1)}, J_-^{(0)}] + \lambda^{-1} [J_{+}^{(0)}, J_-^{(1)}]  \nonumber \\ 
={}& \lambda^{-1} \left( \partial_+ J_-^{(1)} + [J_{+}^{(0)}, J_-^{(1)}] \right) - \lambda \left( \partial_- J_+^{(1)} + [J_-^{(0)},J_{+}^{(1)}] \right) \nonumber \\ 
={}& 0\,,
\end{align}
where we have utilized the on-shell relation (\ref{rel1}) and (\ref{rel2}) in the last steps.

\medskip 

Our goal here is to see whether the machine-learning procedure reconstructs this family from the data obtained from the equation of motion.

\subsection{A linear map from LI current to Lax connection}

Analogous to the PCM case, we consider a linear map defined as
\begin{equation}
L_+=a\,J_+^{(0)}+b\,J_+^{(1)}\,,
\qquad
L_-=c\,J_-^{(0)}+d\,J_-^{(1)}\,,
\label{eq:sym-abcd-ansatz}
\end{equation}
where $a,b,c,d\in\mathbb C$\,. This setup serves as the minimal ansatz for the associated Lax connection in our machine-learning training procedure. 

\medskip

Due to the flatness of the Lax connection, the parameters are not independent. Indeed, by substituting the map (\ref{eq:sym-abcd-ansatz}), one can evaluate the flatness condition as follows: 
\begin{eqnarray}
  0 &=& \partial_+ L_- - \partial_- L_+ + [L_+,L_-] \nonumber \\ 
  &=& c \,\partial_+ J_-^{(0)} - a\, \partial_- J_+^{(0)} 
  + ac\, [J_+^{(0)},J_-^{(0)}] + bd\, [J_+^{(1)}, J_-^{(1)}] \nonumber \\ 
&& + d\, \partial_+ J_-^{(1)} -b\,\partial_- J_+^{(1)} + ad [J_+^{(0)}, J_-^{(1)}] - bc [J_-^{(0)}, J_+^{(1)}]\,. \label{eq:sym-flatness-abcd}
\end{eqnarray}
The grade-zero part is removed by using \eqref{scsm-flat1}, which gives
$a=c$ and $a=c=ac=bd$.  The possible branches are $a=c=1$ with $bd=1$,
and $a=c=0$ with $bd=0$.  However, the grade-one part must vanish by using
\eqref{rel1} and \eqref{rel2}.  This excludes the latter branch, and the
solution is given by
\begin{equation}
a=1\,,
\qquad
c=1\,,
\qquad
bd=1\,.
\label{eq:sym-expected}
\end{equation}
Equivalently, by introducing
\begin{equation}
b=\lambda,
\qquad
d=\lambda^{-1},
\label{eq:sym-lambda-param}
\end{equation}
the standard one-parameter family \eqref{eq:sym-standard-lax} is recovered.

\medskip 

In the following, we will try to reproduce the parameter relations by using the machine-learning procedure. In fact, we will see that the trained coefficients are plotted on the curves specified by \eqref{eq:sym-expected}.

\subsection{Machine-learning setup}

To understand what relations should emerge from training, it is useful to consider the simplest symmetric coset,
\begin{equation}
S^2 \simeq SU(2)/U(1).
\end{equation}

\medskip 
\paragraph{Linear ansatz.}
Following the parametrization in \eqref{eq:sym-abcd-ansatz}, we employ the following linear ansatz for the Lax connection components:
\begin{align}
    L_+(J_+^{(0)}, J_+^{(1)}; a, b) &= a\,J_+^{(0)}+b\,J_+^{(1)}\,, \\
    L_-(J_-^{(0)}, J_-^{(1)}; c, d) &= c\,J_-^{(0)}+d\,J_-^{(1)}\,,
\end{align}
where $a, b, c, d\in\mathbb C$ are constant coefficients to be determined through the optimization process.
In the numerical results presented later, we employ the real slice $a, b, c, d \in \mathbb{R}$ as in the case of the PCM.

\paragraph{Data generation and sampling strategy.}
Training data are generated at the level of local current components.  For
each sample we draw
\begin{equation}
J_+^{(0)},\quad J_-^{(0)},\quad
J_+^{(1)},\quad J_-^{(1)},
\label{eq:sym-sampling}
\end{equation}
together with one grade-zero derivative, which we take to be
$\partial_-J_+^{(0)}$.  The remaining derivatives are then fixed by the
on-shell relations written above.  Solving \eqref{rel1} and \eqref{rel2}
for the grade-one derivatives gives
\begin{align}
\partial_+J_-^{(1)}
&=
-[J_+^{(0)},J_-^{(1)}],
\nonumber\\
\partial_-J_+^{(1)}
&=
-[J_-^{(0)},J_+^{(1)}],
\label{eq:sym-sampled-grade-one-derivatives}
\end{align}
while the grade-zero condition \eqref{scsm-flat1} fixes
\begin{equation}
\partial_+J_-^{(0)}
=
\partial_-J_+^{(0)}
-[J_+^{(0)},J_-^{(0)}]
-[J_+^{(1)},J_-^{(1)}].
\label{eq:sym-sampled-grade-zero-derivative}
\end{equation}
Thus every sampled point satisfies the equations of motion and the
off-shell flatness condition.  The extra sampling of
$\partial_-J_+^{(0)}$ is needed because the grade-zero condition determines only one relation between
$\partial_+J_-^{(0)}$ and $\partial_-J_+^{(0)}$.  If one imposed the
analytic expectation $a=c$ from the outset, only the difference of these
two derivatives would enter the scalar ansatz and this additional sampling
would be unnecessary.  In the learning experiment, however, $a$ and $c$
are treated as independent parameters, so both derivatives must be
available.

\paragraph{Loss function.}
The principal training objective is the on-shell flatness of the candidate
connection.  For the symmetric-coset ansatz, the curvature residual is
\begin{align}
F_{+-}(L)
&=
\partial_+L_- - \partial_-L_+ + [L_+,L_-] \notag \\
&=
c\,\partial_+J_-^{(0)}
-a\,\partial_-J_+^{(0)}
+ac\,[J_+^{(0)},J_-^{(0)}]
+bd\,[J_+^{(1)},J_-^{(1)}]
\nonumber\\
&\quad+
d\,\partial_+J_-^{(1)}
-b\,\partial_-J_+^{(1)}
+ad\,[J_+^{(0)},J_-^{(1)}]
-bc\,[J_-^{(0)},J_+^{(1)}].
\label{eq:sym-flatness-loss-residual}
\end{align}
The derivatives appearing in this expression are not independent training
variables.  They are fixed from the sampled current data
\eqref{eq:sym-sampling} by the on-shell substitutions
\eqref{eq:sym-sampled-grade-one-derivatives} and
\eqref{eq:sym-sampled-grade-zero-derivative}.  
Given a batch of $N$ samples, we define the flatness loss as
\begin{equation}
\mathcal L_{\rm flat}
=
\frac{1}{N}\sum_{n=1}^{N}
\left\|
F_{+-}^{(n)}(L)
\right\|^2
=
\frac{1}{N}\sum_{n=1}^{N}
\left\|
\partial_+L_-^{(n)}
-\partial_-L_+^{(n)}
+[L_+^{(n)},L_-^{(n)}]
\right\|^2,
\label{eq:sym-flatness-loss}
\end{equation}
where the norm is defined in \eqref{eq:matrix-norm}.
Thus the optimization minimizes the on-shell curvature of the candidate
connection.

\paragraph{Regularization against trivial collapse.}
In addition to the flatness loss, we include a mild auxiliary
regularization term.  
This is used to suppress degenerate solutions in which the flatness loss is made small by collapsing some coefficients, for example by driving the coset coefficients toward $b=d=0$.  
The total training objective is
\begin{equation}
\mathcal L_{\rm total}
=\mathcal L_{\rm flat}+\mathcal L_{\rm reg},
\end{equation}
where
\begin{align}
\mathcal L_{\rm reg}
=&
w_{\rm nz}\,\frac{1}{4}\sum_{\theta\in\{a,b,c,d\}}
\exp\!\left(-\gamma\,\theta^2\right)
\nonumber\\
&+
w_{\rm margin}\,\frac{1}{4}\sum_{\theta\in\{a,b,c,d\}}
\left[\max\!\left(0,m-|\theta|\right)\right]^2
\nonumber\\
&+
w_{\rm coll}
\left[
\max\!\left(
0,\,
r-\sqrt{\frac{1}{4}\sum_{\theta\in\{a,b,c,d\}}\theta^2+\epsilon}
\right)
\right]^2 .
\label{eq:s2-regularization}
\end{align}
Here, $\gamma, m,r,\epsilon$ are real hyper parameters and $w_\text{nz}, w_\text{margin}, w_\text{coll}$ are the weights of the three terms. 
The first two terms keep the individual coefficients away from the
near-zero region, while the third term penalizes a collective collapse of
all four coefficients.  
In the numerical runs shown below we use
\begin{equation}
\begin{aligned}
&w_{\rm nz}=0.6,\qquad
\gamma=10.0,\qquad
w_{\rm margin}=1.0,\qquad
m=0.9,\\
&w_{\rm coll}=1.2,\qquad
r=0.95,\qquad
\epsilon=10^{-12}.
\end{aligned}
\end{equation}

\subsection{Visualization of learned results}

\begin{figure}[t]
\centering

\begin{subfigure}[t]{0.48\textwidth}
  \centering
  \includegraphics[width=\textwidth]{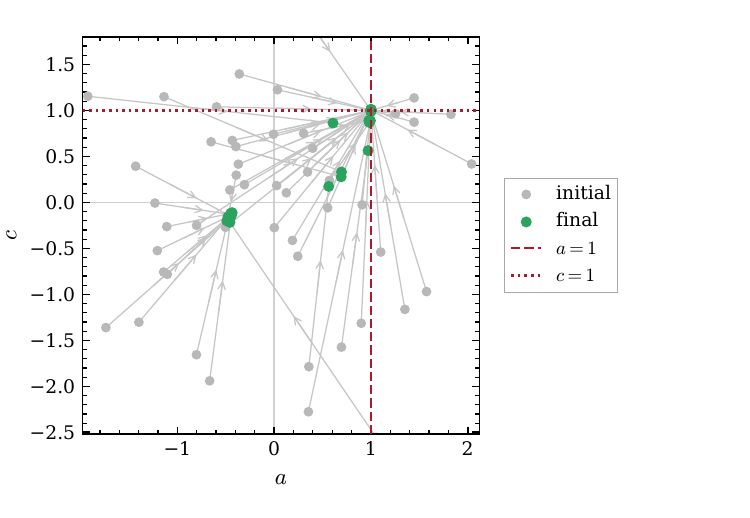}
  \caption{
  \footnotesize
  Learned coefficients $(a,c)$. Gray points denote initial values and
  green points final values. The dashed lines indicate the analytic expectation
  $a=1$ and $c=1$.}
  \label{fig:s2-ac}
\end{subfigure}
\hfill
\begin{subfigure}[t]{0.48\textwidth}
  \centering
  \includegraphics[width=\textwidth]{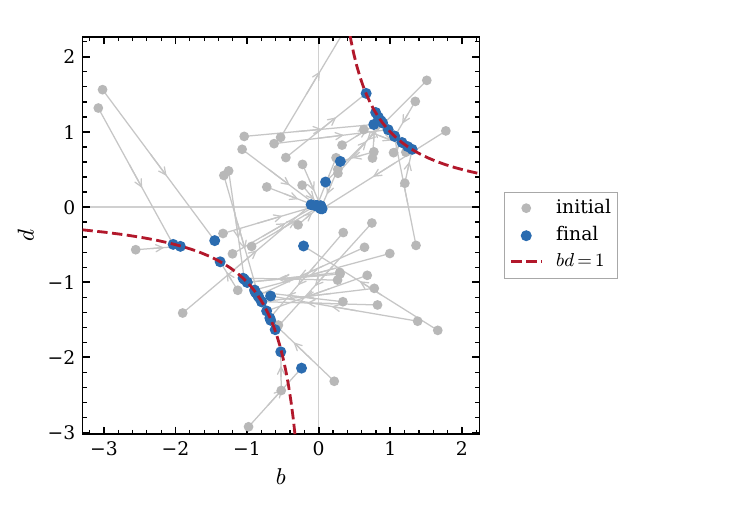}
  \caption{
  \footnotesize
  Learned coefficients $(b,d)$. Gray points denote initial values and
  blue points final values. The dashed curve is the analytic expectation
  $bd=1$.}
  \label{fig:s2-bd}
\end{subfigure}

\caption{Scatter plots of the learned coefficients for the symmetric-coset
ansatz \eqref{eq:sym-abcd-ansatz}.}
\label{fig:s2-coeff-scatter}
\end{figure}

The learning results are summarized in
Fig.~\ref{fig:s2-coeff-scatter}, whose two panels probe different
parameter subspaces of the ansatz \eqref{eq:sym-abcd-ansatz}.

\medskip

Figure~\ref{fig:s2-ac} shows the distribution of the coefficients $(a,c)$ associated with the $\mathfrak{h}$-part. Except for a few runs, a substantial subset of runs is attracted toward the analytic expectation $a=1$ and $c=1$ (dashed lines). This provides evidence that the model correctly tends to recover the canonical normalization of the gauge part.

\medskip

The behavior of the $\mathfrak{m}$-part coefficients $(b,d)$, shown in Fig.~\ref{fig:s2-bd}, exhibits a much clearer pattern. The final points are strongly attracted toward the hyperbola $bd=1$ (dashed curve). This relation is the defining feature of the one-parameter family \eqref{eq:sym-lambda-param}.

\medskip 

Crucially, these results demonstrate that the machine-learning procedure does not converge to a single isolated flat connection. Rather, it successfully reconstructs the continuous one-parameter locus indicative of a genuine Lax family, with the spectral parameter emerging naturally as the degree of freedom along the $bd=1$ curve.

\medskip 

There is a remark regarding the exceptional points observed in the
optimization.  In Fig.~\ref{fig:s2-ac}, some runs do not converge toward
$(a,c)=(1,1)$.  Examining the corresponding coefficients in
Fig.~\ref{fig:s2-bd} shows that these runs are located close to
$(b,d)=(0,0)$.  This is the same collapse direction that motivates the
auxiliary regularization above.  Indeed, when $b$ and $d$ are both small,
the terms involving the coset part of the candidate connection are strongly
suppressed in the curvature residual \eqref{eq:sym-flatness-loss-residual}.
In particular, the term $bd[J_+^{(1)},J_-^{(1)}]$ cannot cancel the
grade-zero condition.  Hence this branch cannot make the
flatness loss vanish.

\medskip

The regularization penalizes this collapse direction, so these points are
not competitive minima of the total objective.  Rather, they should be
viewed as high-loss local minima reached from a subset of initializations.
Numerically, their final flatness losses are much larger than those of the
runs that converge to the $bd=1$ curve.

\subsection{Explicit learning of the spectral parameter}
\label{sec:priorfree-body}

The experiments above learn $a,b,c,d$ as \emph{constants} and record where they land on the
curve~\eqref{eq:sym-expected}: they return points of the family, not the dependence on the
spectral parameter $\lambda$ itself. As in the PCM (Sec.~\ref{sec:pcm-lambda-body}) the next step is to
promote the coefficients to \emph{functions} of an explicit input $\lambda$. We do so for all four at once
in a truncated Laurent ansatz
\begin{equation}
  f\,(\lambda)=\sum_{k=-K}^{K}f_k\,\lambda^{k} ,
  \quad f\in\{a,b,c,d\},
  \label{eq:s2-laurent}
\end{equation}
where $K$ is a cutoff order, taken to be $K=2$ for simplicity\footnote{%
  A larger $K$ would be preferable, so that the ansatz 
  can express a wide range of functions.
  The Laurent basis on the annulus does not permit it: across
  $|\lambda|\in[0.25,4]$ the extreme modes $\lambda^{\pm K}$ differ in magnitude by
  $4^{2K}$, so at high order the residual and its gradient are dominated by those modes at
  initialization and training diverges. We therefore work at the low end of the usable
  range.
  Whether this truncates the result can be checked after training: see Appendix~\ref{app:s2-setup} for details.
  },
and
$\lambda$ sampled on an
origin-centred annulus $|\lambda|\in[0.25,4]$
to avoid possible pole at $\lambda=0$.
We use the explicit Laurent ansatz here rather than the neural-network
ansatz of Section~\ref{sec:pcm-lambda-body} because the coefficients of a
Lax connection are meromorphic functions of $\lambda$, and the standard
$S^2$ coefficients are represented exactly by Laurent monomials.
The explicit basis also makes the learned $\lambda$-dependence directly
readable mode by mode.
Training minimizes the scale-invariant component-norm residual
$\mathcal{L}_{\rm cn}$ defined in \eqref{eq:pcm-loss-cn}, now averaged over
both $\lambda$ and the on-shell currents.

\medskip

The normalized residual removes the incentive to collapse onto the trivial
connection $L=0$
as we noted in Appendix~\ref{app:pcm-lambda},
but it does not forbid a $\lambda$-\emph{independent} one: the constant
chart $b(\lambda)= \text{const}.$, with $d=1/b$ and $a=c=1$, is a genuine non-zero point of
\eqref{eq:sym-expected} and hence exactly flat.
Minimizing flatness alone
therefore collapses onto it, as the control run of Appendix~\ref{app:priorfree-lambda} shows.
In the PCM case, this problem was resolved by prescribing one coefficient
explicitly as $a(\lambda)=\lambda$.
Here we do not introduce such a parameter fixing since that presupposes which coefficients carry the parameter, which are $b$ and $d$ in the present case.
We add instead a term to the loss function that forbids the collapse to constant coefficients given by
\begin{equation}
  r_f=\frac{|f_{0}|^{2}+\varepsilon}{\sum_{k}|f_{k}|^{2}+\varepsilon}\in[0,1],
  \qquad f\in\{a,b,c,d\},
  \label{eq:nc-fraction}
\end{equation}
for the fraction of a coefficient carried by its constant mode; it is equal to $1$ precisely when $f$ is constant, including when $f\equiv0$. Then, we form the total loss function as
\begin{equation}
  \mathcal{L}_{\rm cn}
  +w_{\rm nc}  \prod_{f\in\{a,b,c,d\}}\!\! r_f
  +w_{\rm reg} \sum_{f}\sum_{|k|\ge2} |f_k|^{2}\,.
  \label{eq:nc-penalty}
\end{equation}
The $w_{\rm nc}$ term is maximal exactly on the $\lambda$-independent charts (and on $L=0$), and is
suppressed as soon as any one coefficient acquires non-constant content.
Being symmetric in $a,b,c,d$ and sensitive only to $|f_0|$ against the rest, it penalizes
constancy alone and leaves every other feature of the coefficients free.
The $w_{\rm reg}$ term penalizes the modes with $|k|\ge2$. Flatness alone
leaves the degree of the chart undetermined, since $bd=1$ holds for inverse monomials of
any degree, and this term selects the lowest one.

\medskip

With the loss function \eqref{eq:nc-penalty}, the correct solution is recovered in every run
in which the flatness residual is successfully minimized:\footnote{%
  The minimization does not succeed from every initialization. About half of the runs stall
  short of the family, for a reason traced in Appendix~\ref{app:s2-results}.}
$a$ and $c$ automatically stay pinned at unity to $10^{-5}$ while $b$ and $d$ move
onto single, mutually inverse Laurent modes satisfying $bd=1$ (Table~\ref{tab:priorfree-learned}).
The learned chart is
\begin{equation}
  a(\lambda)=c(\lambda)=1,\qquad
  b(\lambda)=e^{i\varphi}\lambda^{\,n},\qquad
  d(\lambda)=e^{-i\varphi}\lambda^{-n}\,.
  \label{eq:priorfree-learned}
\end{equation}
The degree $|n|=1$, the phase $\varphi$ and the sign of $n$ are all learned, not imposed.
In the best of the runs, $bd=1$ holds to $2\times10^{-7}$,
the flatness residual is $3\times10^{-15}$, and $\arg b$ winds once around $\lambda=0$.
Appendix~\ref{app:priorfree-lambda} collects the diagnostics behind these numbers and the
per-seed results.

\medskip

The amplitude, the phase $\varphi$ and the sign of $n$ in \eqref{eq:priorfree-learned} are
gauge. 
Flatness is imposed at each $\lambda$ separately, so the change of
variable
\begin{equation}
\lambda\mapsto \xi\,\lambda^{\pm1},
\qquad
\xi\in\mathbb C^\times=\mathbb{C}\setminus\{0\},
\end{equation}
leaves the loss invariant while changing all three quantities.
This is why the runs reported here land on $n=-1$ rather than on the standard $n=+1$ of
\eqref{eq:sym-standard-lax}.
What is \emph{not} gauge is that $b$ and $d$ come out as
single monomials. For finite Laurent polynomials, this follows from the
exact relation $bd=1$, whose violation is measured by $|bd-1|$.

\begin{table}[htbp]
\centering
\caption{Laurent mode moduli $|f_k|$ of the coefficients learned with
  \eqref{eq:nc-penalty} on the origin-centred annulus $|\lambda|\in[0.25,4]$ with $K=2$:
  $b$ becomes a pure $k=-1$ monomial and $d$ its inverse.
  $a$ and $c$ sit on their $k=0$ mode with modulus $1.000$ and are omitted.
  Only moduli are shown; the phases are gauge.
  It is the best of the runs reported in Appendix~\ref{app:priorfree-lambda}.}
\label{tab:priorfree-learned}
\begin{tabular}{lccccc}
\hline
 & $k{=}{-}2$ & $k{=}{-}1$ & $k{=}0$ & $k{=}{+}1$ & $k{=}{+}2$ \\
\hline
$b$ & $4\times10^{-4}$ & $\mathbf{1.006}$ & $5\times10^{-4}$ & $<\!10^{-4}$ & $<\!10^{-4}$ \\
$d$ & $<\!10^{-4}$ & $<\!10^{-4}$ & $3\times10^{-4}$ & $\mathbf{0.994}$ & $5\times10^{-4}$ \\
\hline
\end{tabular}
\end{table}

\medskip

Two limitations are worth noting.
The qualitative outcome ($a=c=1$ with two inverse monomials of
winding one) is common to all successful runs, but the accuracy is not: they span six
orders of magnitude in $\mathcal{L}_{\rm cn}$ (Table~\ref{tab:annulus-modes} in Appendix~\ref{app:priorfree-lambda}).
 The spread
reflects slow convergence, which a better training strategy should reduce.
Another is the ansatz itself. Expanded about the origin, the Laurent form
pins the poles of the coefficients to $\lambda=0,\infty$,
which suffices for $S^2$ but not for the PCM, whose coefficients
\eqref{eq:pcm-ac-lambda} are singular at $\lambda=\pm1$. Learning the pole locations, with a
rational ansatz $b=P(\lambda)/Q(\lambda)$, is the natural next step.
We leave such improvements to future work.

\section{A test for non-symmetric coset \texorpdfstring{$T^{1,1}$}{T1,1}}
\label{sec:t11-field}

Let us next apply the previously developed machine-learning technique to a
non-symmetric coset, $T^{1,1}$. This case provides a useful test of the
approach, since $T^{1,1}$ can be formulated as a coset space and the same
procedure remains applicable in principle.

\medskip 

In \cite{Krippendorf:2021lee}, non-integrable 
systems were investigated using a machine-learning framework, where it was reported that the loss function failed to converge. In contrast, we will demonstrate below that the loss function can indeed converge even within non-integrable systems. This unexpected behavior originates from the so-called ``fake Lax'' problem inherent in the study of integrable systems. Consequently, our findings indicate that the machine-learning procedure may inadvertently capture a fake Lax pair, suggesting that utmost caution is required when searching for Lax connections via machine-learning techniques.

\subsection{Coset structure of \texorpdfstring{$T^{1,1}$}{T1,1}}

It is known that $T^{1,1}$ is a five-dimensional Sasaki-Einstein manifold, which is known as the base manifold of the conifold \cite{Candelas:1989js}. That is, the conifold is realized as a cone over $T^{1,1}$\,. The metric of $T^{1,1}$ is given by 
\begin{equation}
   ds^2_{T^{1,1}}
=
\frac16\sum_{i=1}^2
\left(d\theta_i^2+\sin^2\theta_i\,d\phi_i^2\right)
+\frac19
\left(d\psi+\cos\theta_1\,d\phi_1+\cos\theta_2\,d\phi_2\right)^2\,.
\label{eq:t11-metric}
\end{equation}

\medskip 

The $T^{1,1}$ geometry can be described as a coset space, and its metric can be derived via the coset construction, analogous to the cases of (anti-)de Sitter spaces and round spheres. For this purpose, following \cite{Crichigno:2014ipa}, we represent $T^{1,1}$ as the coset space $M=G/H$:
\begin{equation}
T^{1,1}
=
\frac{SU(2)_1\times SU(2)_2\times U(1)_R}{U(1)_1\times U(1)_2}\,. 
\label{eq:t11-coset}
\end{equation}
Here $G=SU(2)_1\times SU(2)_2\times U(1)_R$ and $H=U(1)_1\times U(1)_2$\,. Let $K_i$ and $L_i$ ($i=1,2,3$) be the generators of $SU(2)_1$ and $SU(2)_2$, respectively, and let $M$ denote the generator of $U(1)_R$. These generators satisfy the following commutation relations:
\begin{equation}
[K_i,K_j]=\epsilon_{ij}{}^kK_k,
\quad
[L_i,L_j]=\epsilon_{ij}{}^kL_k,
\quad
[K_i,L_j]=0,
\quad
[M,K_i]=[M,L_i]=0.
\end{equation}
The Lie algebra $\mathfrak{h}$ of the denominator subgroup $H$ is spanned by $T_1$ and $T_2$, which are defined as
\begin{equation}
T_1:=K_3+L_3,
\qquad
T_2:=K_3-L_3+4M,
\end{equation}
while the remaining singlet direction is given by
\begin{equation}
T:=K_3-L_3+M.
\end{equation}
Consequently, the Lie algebra $\mathfrak{g}$ of $G$ is decomposed as a direct sum of vector spaces,  
\begin{equation}
\mathfrak g=\mathfrak h\oplus\mathfrak m,
\qquad
\mathfrak h=\mathrm{span}_{\mathbb{R}}\{T_1,T_2\},
\qquad
\mathfrak m=\mathrm{span}_{\mathbb{R}}\{K_1,K_2,L_1,L_2,T\},
\label{eq:t11-hm-decomposition}
\end{equation}
where $\mathfrak{m}$ is the orthogonal complement to $\mathfrak{h}$. 

\medskip 

The structure of the $T^{1,1}$ coset \eqref{eq:t11-coset} is reflected in
the adjoint action of $\mathfrak h$ on $\mathfrak m$:
\begin{align}
[T_1,K_1]&=K_2, & [T_1,K_2]&=-K_1, \nonumber 
&
[T_2,K_1]&=K_2, & [T_2,K_2]&=-K_1, \nonumber \\
[T_1,L_1]&=L_2, & [T_1,L_2]&=-L_1, \nonumber
&
[T_2,L_1]&=-L_2, & [T_2,L_2]&=L_1,\\
[T_1,T]&=0, & [T_2,T]&=0.
\label{eq:t11-commutators}
\end{align}
Furthermore, since $K_3$ and $L_3$ are expressed as
\begin{equation}
K_3=\frac12T_1-\frac16T_2+\frac23 T,
\qquad
L_3=\frac12T_1+\frac16T_2-\frac23 T,
\end{equation}
the $SU(2)$ commutation relations yield
\begin{equation}
[K_1,K_2] = \frac12T_1-\frac16T_2+\frac23 T,
\qquad
[L_1,L_2] = \frac12T_1+\frac16T_2-\frac23 T.
\label{eq:t11-nonsymmetric-commutators}
\end{equation}
Thus the commutators of two coset generators have both $\mathfrak h$- and
$\mathfrak m$-components.  In total, these relations immediately imply that
the $T^{1,1}$ coset is reductive but non-symmetric, i.e.,
\begin{equation}[\mathfrak h,\mathfrak m]\subset\mathfrak m,\qquad[\mathfrak m,\mathfrak m]\not\subset\mathfrak h.
\end{equation}
This non-symmetric property poses the algebraic obstruction to applying the standard symmetric-coset Lax connection without modification.

\subsubsection*{Coordinate Realization and Metric}

For later convenience, let us introduce the coordinates concretely by taking the group parametrization as  
\begin{eqnarray}
   g =  \exp\left(
\phi_1 K_3 + \phi_2 L_3 - 2\psi M \right)
\exp\left(
\theta_1 K_2 + (\theta_2 + \pi) L_2 
\right)\,. 
\label{para}
\end{eqnarray}

For a coordinate realization, we take
\begin{equation}
(\theta_1,\phi_1,\theta_2,\phi_2,\psi)
\end{equation}
as target-space coordinates. With a conventional choice of coset
representative, the coset part of the current can be written as
\begin{equation}
J_\pm^{(1)}
=
-\sin\theta_1\,\partial_\pm\phi_1\,K_1
+\partial_\pm\theta_1\,K_2
+\sin\theta_2\,\partial_\pm\phi_2\,L_1
+\partial_\pm\theta_2\,L_2
+\frac23\eta_\pm\, T,
\label{eq:t11-coordinate-current}
\end{equation}
where
\begin{equation}
\eta_\pm
:=
\partial_\pm\psi
+\cos\theta_1\,\partial_\pm\phi_1
+\cos\theta_2\,\partial_\pm\phi_2.
\end{equation}
Using the supertrace, the pullback of the target-space metric to the
world sheet is
\begin{align}
-\frac13\operatorname{Str}
\left(J_+^{(1)}J_-^{(1)}\right)
&=
\frac16\sum_{i=1}^{2}
\left(
\partial_+\theta_i\,\partial_-\theta_i
+\sin^2\theta_i\,
\partial_+\phi_i\,\partial_-\phi_i
\right)
+\frac19\eta_+\eta_-.
\label{eq:t11-lagrangian-density}
\end{align}
This is the sigma-model kinetic density associated with the
target-space metric \eqref{eq:t11-metric}.

\subsection{Current Equations for the Non-Symmetric Coset}

With the same normalization as in \eqref{eq:t11-lagrangian-density}, the
two-dimensional sigma-model action can be written as
\begin{equation}
S
=\frac12
\int d\sigma^+d\sigma^-\,
\left(
-\frac13\,\operatorname{Str}(J_+^{(1)}J_-^{(1)})
\right).
\end{equation}
We now state the current equations in the notation used below. Decompose
\begin{equation}
J_\pm=g^{-1}\partial_\pm g=A_\pm+K_\pm,
\qquad
A_\pm:=J_\pm^{(0)}\in\mathfrak h,
\qquad
K_\pm:=J_\pm^{(1)}\in\mathfrak m.
\end{equation}
The off-shell flatness conditions for $J_\pm$ give
\begin{align}
\partial_+A_- - \partial_-A_+ + [A_+,A_-]
+P_{\mathfrak h}([K_+,K_-])&=0,
\label{eq:t11-h-mc}\\
\partial_+K_- - \partial_-K_+
+[A_+,K_-]-[A_-,K_+]
+P_{\mathfrak m}([K_+,K_-])&=0.
\label{eq:t11-m-mc}
\end{align}
The sigma-model equation of motion is
\begin{equation}
\partial_+K_-+\partial_-K_+
+[A_+,K_-]+[A_-,K_+]=0.
\label{eq:t11-current-eom}
\end{equation}
If we define
\begin{equation}
C_{\mathfrak h}:=P_{\mathfrak h}([K_+,K_-]),
\qquad
C_{\mathfrak m}:=P_{\mathfrak m}([K_+,K_-]),
\end{equation}
then \eqref{eq:t11-h-mc}--\eqref{eq:t11-current-eom} imply
\begin{equation}
F_{+-}(A)
:=
\partial_+A_- - \partial_-A_+ + [A_+,A_-]
=-C_{\mathfrak h},
\label{eq:t11-a-curvature}
\end{equation}
and
\begin{equation}
\partial_+K_-+[A_+,K_-]=-\frac12C_{\mathfrak m},
\qquad
\partial_-K_+ +[A_-,K_+]=+\frac12C_{\mathfrak m}.
\label{eq:t11-cross-derivatives-current}
\end{equation}
The additional $C_{\mathfrak m}$ terms in the two equations of
\eqref{eq:t11-cross-derivatives-current} are absent for symmetric
cosets. They are the current-level manifestation of the non-symmetric
commutators \eqref{eq:t11-nonsymmetric-commutators}, and they are
precisely the terms that the learning ansatz must accommodate.

\medskip

It is convenient to introduce the $H$-covariant derivative acting on
$\mathfrak m$-valued currents by
\begin{equation}
D_\mu
\coloneqq 
\partial_\mu  + [A_\mu,\phantom{K}].
\label{eq:t11-covariant-derivative}
\end{equation}
With this notation, the $\mathfrak m$-component of the off-shell flatness \eqref{eq:t11-m-mc} and the equation of motion \eqref{eq:t11-current-eom} take the compact form
\begin{align}
D_+K_- - D_-K_+ + C_{\mathfrak m}=0,
\qquad
D_+K_- + D_-K_+=0.
\label{eq:t11-covariant-mc-eom}
\end{align}

\subsection{Collapse of the minimal ansatz}

We now make precise why the standard symmetric-coset Lax ansatz does not yield a nontrivial spectral-parameter family for $T^{1,1}$.  Consider the direct analogue of the symmetric-space ansatz,
\begin{equation}
L_+ = A_+ + uK_+,
\qquad
L_- = A_- + vK_-,
\label{eq:t11-uv-ansatz}
\end{equation}
where $u,v$ are constant complex parameters.  Using
\begin{equation}
F_{+-}(A)=-C_{\mathfrak h},
\qquad
D_+K_-=-\frac12 C_{\mathfrak m},
\qquad
D_-K_+=+\frac12 C_{\mathfrak m},
\end{equation}
one obtains
\begin{align}
F_{+-}(L)
&=
F_{+-}(A)
+ vD_+K_-
- uD_-K_+
+ uv[K_+,K_-]
\nonumber\\
&=
(uv-1)C_{\mathfrak h}
+
\left(
uv-\frac{u+v}{2}
\right)C_{\mathfrak m}.
\label{eq:t11-uv-flatness}
\end{align}
For a symmetric coset, $C_{\mathfrak m}=0$, and flatness imposes only the single condition $uv=1$.  This leaves a nontrivial one-parameter family, for instance
\begin{equation}
u=\lambda,\qquad v=\lambda^{-1},
\end{equation}
which is the usual spectral-parameter dependence.

\medskip

For $T^{1,1}$, by contrast, the commutation relations in
\eqref{eq:t11-nonsymmetric-commutators} imply that $C_{\mathfrak m}$ does not vanish identically.  Requiring the flatness of
\eqref{eq:t11-uv-ansatz} for generic on-shell currents therefore gives the two independent conditions
\begin{equation}
uv=1,
\qquad
uv-\frac{u+v}{2}=0.
\end{equation}
Equivalently,
\begin{equation}
uv=1,
\qquad
u+v=2.
\end{equation}
Thus $u$ and $v$ are the two roots of
\begin{equation}
(z-1)^2=0,
\end{equation}
and the only solution is
\begin{equation}
u=v=1.
\end{equation}

\medskip

Hence the one-parameter symmetric-coset Lax family collapses in the non-symmetric $T^{1,1}$ case.  The obstruction is not the absence of flat connections altogether: at $u=v=1$, one simply recovers $L_\pm=A_\pm+K_\pm=J_\pm$, whose flatness is automatically satisfied.  Rather, the obstruction is the additional $C_{\mathfrak m}$ contribution generated by
\begin{equation}
[\mathfrak m,\mathfrak m]\not\subset \mathfrak h .
\end{equation}
This extra term imposes a second algebraic constraint on $u$ and $v$, thereby removing the spectral parameter within the minimal ansatz.

\subsection{Learning Setup for the \texorpdfstring{$T^{1,1}$}{T1,1} Sigma Model}

The analysis above shows that the direct symmetric-coset ansatz
\eqref{eq:t11-uv-ansatz} does not produce a nontrivial spectral-parameter
family for $T^{1,1}$; it collapses to the trivial choice $u=v=1$.
We therefore enlarge the search space.  Instead of imposing the standard
symmetric-space form of the Lax connection, we keep the gauge component
$A_\pm$ in its canonical form and let the machine-learning model determine
how the coset component $K_\pm$ should be transformed.

\medskip

The ansatz below should be understood as a current-level search for
nontrivial flat representatives in an enlarged class.  A genuine
spectral-parameter family is obtained only if the learned maps can be
organized into a continuous family, or if the spectral parameter is included
explicitly as an additional input.

\paragraph{Ansatz.}
We use the current-level ansatz
\begin{equation}
L_+ = A_+ + \ell_+\!\left(K_+\right),
\qquad
L_- = A_- + \ell_-\!\left(K_-\right),
\label{eq:t11-mlp-ansatz}
\end{equation}
where $\ell_\pm:\mathfrak m\to\mathfrak m$ are represented by small
multilayer perceptrons.  Thus the gauge component is kept fixed, while the
coset component is learned from data.

To make this explicit, we use the basis
\begin{equation}
\{T_A\}_{A=1}^{5}
=
\{K_1,K_2,L_1,L_2,T\}
\end{equation}
of $\mathfrak m$.  In this subsection we denote the components of $K_\pm$ by
\begin{equation}
K_\pm
=
k_{1,\pm}K_1+k_{2,\pm}K_2
+l_{1,\pm}L_1+l_{2,\pm}L_2
+t_\pm T.
\label{eq:t11-k-component-expansion}
\end{equation}
Equivalently, in abstract basis notation,
\begin{equation}
K_\pm
=
\sum_{A=1}^{5}K_\pm^A T_A,
\qquad
\ell_\pm(K_\pm)
=
\sum_{A=1}^{5}
\ell_\pm^A(K_\pm^1,\ldots,K_\pm^5)T_A.
\end{equation}
The networks therefore act on the five coefficients $K_\pm^A$.

\paragraph{Loss function.}
The flatness loss is evaluated on sampled current data satisfying the
equations of motion and the off-shell flatness relations:
\begin{equation}
\mathcal L_{\rm flat}
=
\frac{1}{N}\sum_{n=1}^{N}
\left\|
F_{+-}^{(n)}(L)
\right\|^2 =
\frac{1}{N}\sum_{n=1}^{N}
\left\|
\partial_+L_-^{(n)}
-\partial_-L_+^{(n)}
+
[L_+^{(n)},L_-^{(n)}]
\right\|^2,
\end{equation}
where the norm is the same as in \eqref{eq:matrix-norm}.

\medskip

Since $\ell_\pm$ are neural maps, the derivative terms are evaluated by the
chain rule.  For example,
\begin{equation}
\partial_+\ell_-^A(K_-)
=
\sum_{B=1}^{5}
\frac{\partial \ell_-^A}{\partial K_-^B}(K_-)
\,\partial_+K_-^B,
\end{equation}
and similarly for $\partial_-\ell_+^A(K_+)$.  The Jacobian of the neural map is
computed by automatic differentiation, while the derivatives of the input
currents are supplied by the sampled on-shell current data.

\subsection{ML Search for possible Lax connections}

\begin{figure}[t]
\centering
\includegraphics[width=.72\textwidth]{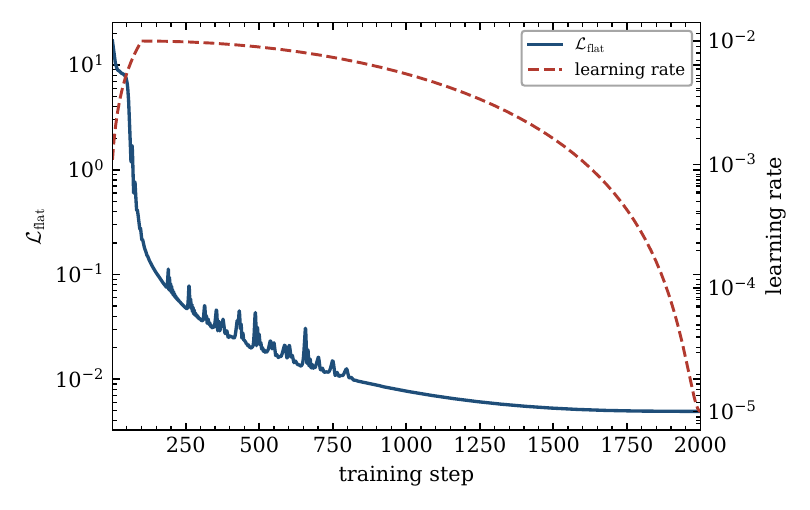}
\caption{Loss and learning rate evolution for the $T^{1,1}$ MLP training run.
The left axis shows the flatness loss, while the right axis shows the learning
rate with warmup followed by cosine annealing.}
\label{fig:t11-loss}
\end{figure}

The training loss decreases by several orders of magnitude in the MLP
experiment, as shown in Fig.~\ref{fig:t11-loss}.  Although this convergence alone is not evidence for the integrability of the full two-dimensional $T^{1,1}$ sigma model, it provides a useful diagnostic: the optimization is attracted to a low-loss region, and thus it is meaningful to inspect the learned maps $\ell_\pm$.

\medskip

\begin{figure}[t]
\centering
\includegraphics[width=\textwidth]{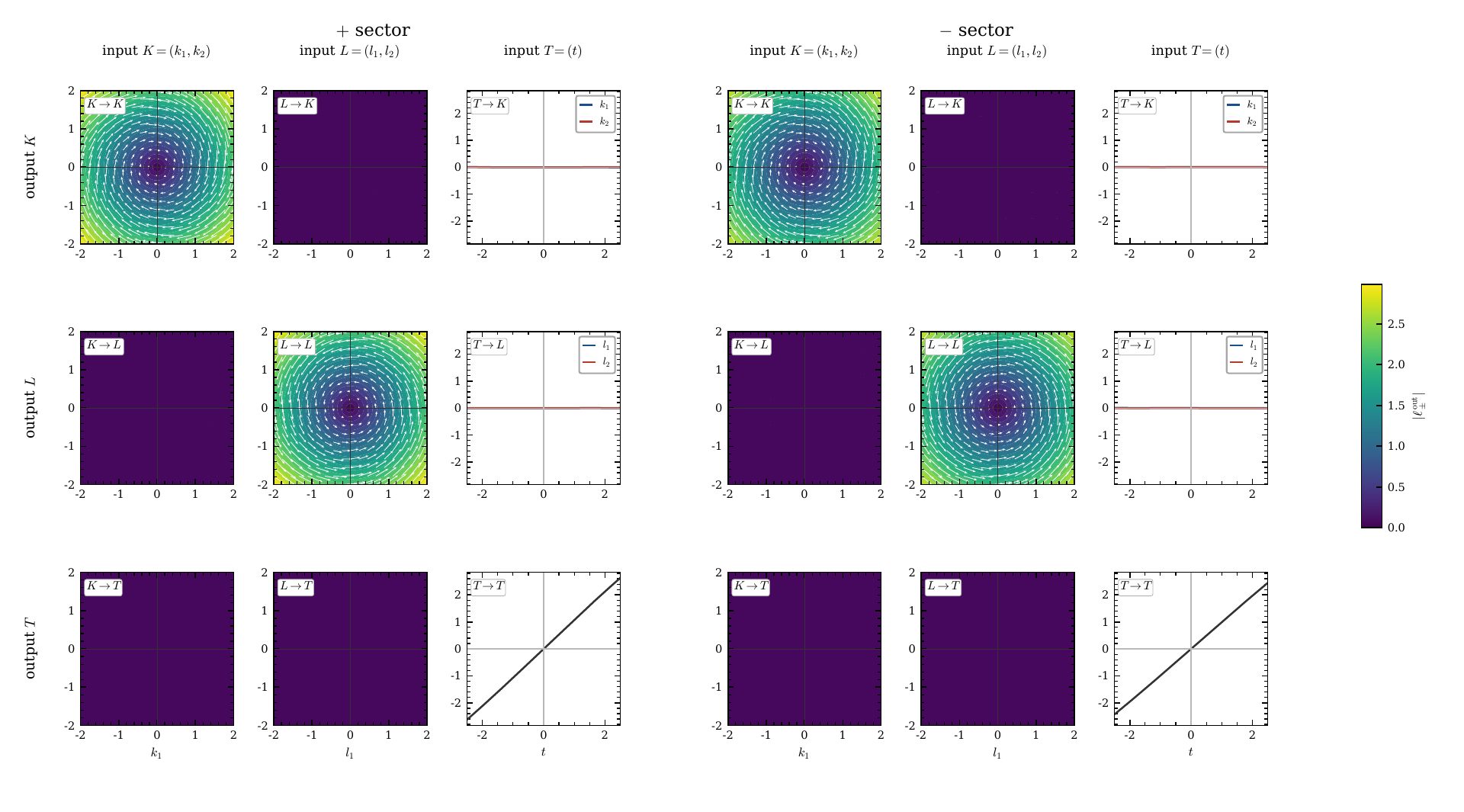}
\caption{Full $3\times 3$ block visualization of the learned maps
$\ell_\pm:\mathbb{R}^5\to\mathbb{R}^5$ for $T^{1,1}$. Rows denote output
blocks $(K,L,T)$ and columns denote input blocks $(K,L,T)$; the $+$ sector is
shown on the left and the $-$ sector on the right. The $K\to K$ and $L\to L$
blocks display dominant rotational vector fields, while $T\to T$ is nearly
one-dimensional and linear. The off-diagonal blocks are much smaller in scale,
supporting the block-diagonal form.}
\label{fig:t11-visualization}
\end{figure}

We therefore visualize the learned maps
$\ell_\pm:\mathbb R^5\to\mathbb R^5$ in the component basis, adapted to the natural $2+2+1$ decomposition
$(K_1,K_2)\oplus(L_1,L_2)\oplus T$ of the coset directions.
Figure~\ref{fig:t11-visualization}
arranges the result as a $3\times 3$ block matrix, with rows corresponding to
output blocks and columns corresponding to input blocks.  The dominant entries
are the $K\to K$ and $L\to L$ vector fields, while the off-diagonal blocks are
much smaller.  Thus the learned maps are far from generic: they are visually
close to block diagonal.

\medskip

The visual pattern in Fig.~\ref{fig:t11-visualization} suggests the following form
\begin{equation}
L_+ = A_+ + \Phi_+\!\left(K_+\right),
\qquad
L_- = A_- + \Phi_-\!\left(K_-\right),
\label{eq:t11-reduced-ansatz}
\end{equation}
where $\Phi_\pm:\mathfrak m\to\mathfrak m$ are constant block-diagonal maps.
Using the component expansion \eqref{eq:t11-k-component-expansion}, write
\begin{equation}
\Phi_\pm(K_\pm)
=
\tilde k_{1,\pm}K_1+\tilde k_{2,\pm}K_2
+\tilde l_{1,\pm}L_1+\tilde l_{2,\pm}L_2
+\tilde t_\pm T.
\end{equation}
The observed structure is captured by
\begin{equation}
\begin{pmatrix}
\tilde k_{1,\pm}\\
\tilde k_{2,\pm}\\
\tilde l_{1,\pm}\\
\tilde l_{2,\pm}\\
\tilde t_\pm
\end{pmatrix}
=
\begin{pmatrix}
a_{K,\pm} & -b_{K,\pm} & 0 & 0 & 0\\
b_{K,\pm} & a_{K,\pm} & 0 & 0 & 0\\
0 & 0 & a_{L,\pm} & -b_{L,\pm} & 0\\
0 & 0 & b_{L,\pm} & a_{L,\pm} & 0\\
0 & 0 & 0 & 0 & c_{T,\pm}
\end{pmatrix}
\begin{pmatrix}
k_{1,\pm}\\
k_{2,\pm}\\
l_{1,\pm}\\
l_{2,\pm}\\
t_\pm
\end{pmatrix}.
\label{eq:t11-rotation-ansatz}
\end{equation}
Thus the $K$- and $L$-blocks are scaled rotations, while the singlet direction
is scaled independently.  In this notation the coefficients $a_{K,L}$ multiply
the identity part of each two-dimensional block, while $b_{K,L}$ multiply the
rotation generator.  This block-diagonal form is not imposed before training;
it is extracted from the structure of the converged neural maps.

\begin{figure}[p]
\centering

\begin{subfigure}[t]{0.95\textwidth}
  \centering
  \includegraphics[width=\textwidth]{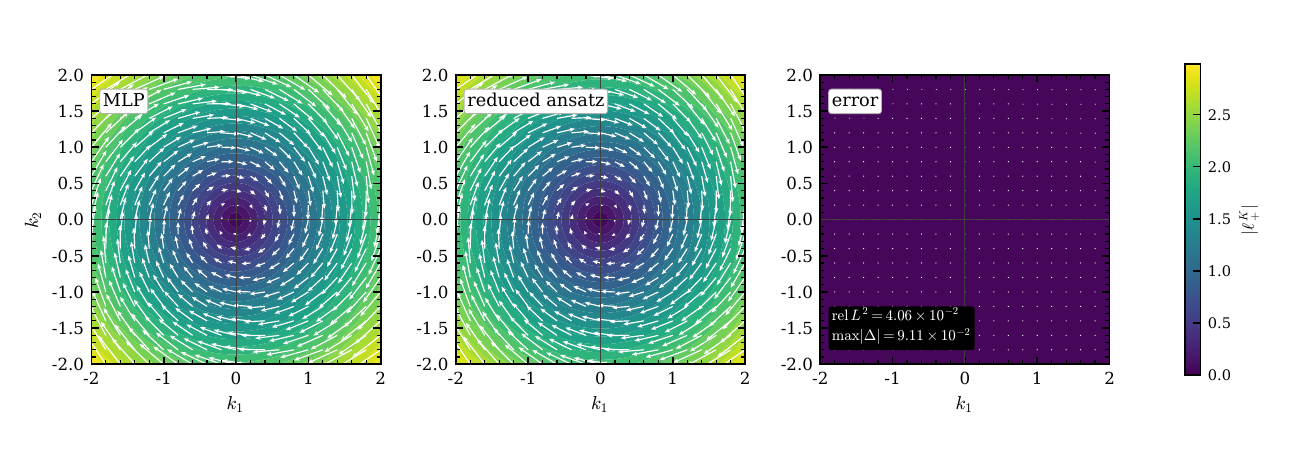}
  \caption{Distillation of the learned $K\to K$ block in the $+$ sector.
  The left panel shows the original MLP map, the middle panel shows the fitted
  block-diagonal form, and the right panel shows the pointwise error.
  The close visual agreement supports the interpretation of the learned
  $K$-block as an almost constant rotation.}
  \label{fig:t11-distill}
\end{subfigure}

\vspace{1.0em}

\begin{subfigure}[t]{0.95\textwidth}
  \centering
  \includegraphics[width=\textwidth]{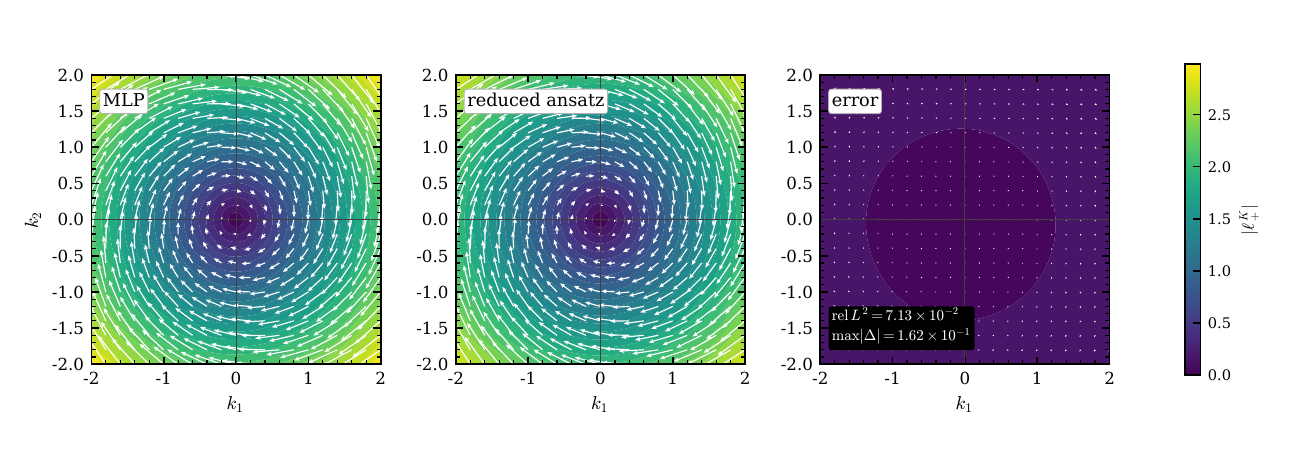}
  \caption{Independent training of the reduced block-diagonal ansatz for the
  $K\to K$ block in the $+$ sector. The reduced model recovers the same
  rotational pattern as the MLP, providing a consistency check of the reduction.}
  \label{fig:t11-independent}
\end{subfigure}

\caption{Comparison of the distilled and independently trained reduced
block-diagonal descriptions of the learned $K\to K$ block in the $+$ sector.}
\label{fig:t11-blockdiag-reduction}
\end{figure}

\paragraph{Quantitative checks}
As a quantitative check of this reduction, we fit the learned map by the
block-diagonal form \eqref{eq:t11-rotation-ansatz}. 
The results are summarized in Fig.~\ref{fig:t11-blockdiag-reduction}.
Figure~\ref{fig:t11-distill}
shows the $K\to K$ block in the $+$ sector.  The fitted block-diagonal rotation
closely reproduces the MLP vector field, with relative $L^2$ error
$4.1\times 10^{-2}$ and maximum absolute error $9.1\times 10^{-2}$ on the
displayed grid.  We also trained the reduced block-diagonal ansatz directly
from scratch; as shown in Fig.~\ref{fig:t11-independent}, the same rotational
pattern is recovered, with a somewhat larger relative $L^2$ error
$7.2\times 10^{-2}$.  These checks support treating
\eqref{eq:t11-rotation-ansatz} as the effective finite-dimensional form of the candidate Lax connection for the analytic calculation below.

\subsection{Analytic Validation of the Reduced Connection}
\label{sec:t11-analytic-validation}

A small flatness loss does not by itself establish that the learned
connection is a genuine Lax connection.  Its flatness condition must encode
the equation of motion, rather than follow solely from the off-shell
flatness identity of the current.  The elementary choice $L_\pm=J_\pm$
illustrates the latter possibility: it is identically flat but contains no
dynamical information.

\medskip

We therefore examine the reduced connection
\begin{equation}
L_+ = A_+ + \Phi_+(K_+),
\qquad
L_- = A_- + \Phi_-(K_-),
\label{eq:t11-validation-connection}
\end{equation}
where $\Phi_\pm$ have the constant block-diagonal form
\eqref{eq:t11-rotation-ansatz}.  Define the equation-of-motion residual by
\begin{equation}
E:=D_+K_-+D_-K_+.
\label{eq:t11-eom-residual}
\end{equation}
Combining this definition with the $\mathfrak m$-component of the off-shell
flatness identity,
\begin{equation}
D_+K_- - D_-K_+ + C_{\mathfrak m}=0,
\end{equation}
gives
\begin{equation}
D_+K_-=\frac12\left(E-C_{\mathfrak m}\right),
\qquad
D_-K_+=\frac12\left(E+C_{\mathfrak m}\right).
\label{eq:t11-derivative-decomposition}
\end{equation}

The block maps commute with the isotropy action, so they may be taken
through the covariant derivatives.  Using
$F_{+-}(A)=-C_{\mathfrak h}$, the curvature becomes
\begin{equation}
\begin{aligned}
F_{+-}(L)
&=
\frac12(\Phi_--\Phi_+)E+R_{\rm alg},
\\
R_{\rm alg}
&:=
[\Phi_+(K_+),\Phi_-(K_-)]
-C_{\mathfrak h}
-\frac12(\Phi_++\Phi_-)C_{\mathfrak m}.
\end{aligned}
\label{eq:t11-flatness-vs-eom}
\end{equation}
Thus, within the present ansatz, flatness can be equivalent to the equation
of motion only if $R_{\rm alg}$ vanishes identically and
$\Phi_--\Phi_+$ has no kernel on the space of admissible
equation-of-motion residuals.

\medskip

We now impose $R_{\rm alg}=0$ for arbitrary local current values.
Let $\Phi_{K,\pm}$ and $\Phi_{L,\pm}$ denote, respectively, the
$K$- and $L$-blocks of the block-diagonal maps $\Phi_\pm$ in
\eqref{eq:t11-rotation-ansatz}.  Matching the coefficients of the
independent current bilinears in the $\mathfrak m$-component of
$R_{\rm alg}$ shows that, on the non-degenerate branch relevant to the
learned maps,
\begin{equation}
c_{T,+}=c_{T,-}=1,
\qquad
\Phi_{K,+}=\Phi_{K,-}=:\Phi_K,
\qquad
\Phi_{L,+}=\Phi_{L,-}=:\Phi_L.
\label{eq:t11-common-blocks}
\end{equation}
The remaining $\mathfrak h$-component requires the transformed
commutators in the $K$- and $L$-planes to reproduce the corresponding
terms in $C_{\mathfrak h}$.  For the two-dimensional blocks in
\eqref{eq:t11-rotation-ansatz}, this condition is equivalent to
\begin{equation}
\det\Phi_K=1,
\qquad
\det\Phi_L=1.
\label{eq:t11-determinant-conditions}
\end{equation}
Consequently, exact cancellation of the algebraic remainder requires
\begin{equation}
\begin{gathered}
c_{T,+}=c_{T,-}=1,
\\
\Phi_{K,+}=\Phi_{K,-}=:\Phi_K,
\qquad
\det\Phi_K=1,
\\
\Phi_{L,+}=\Phi_{L,-}=:\Phi_L,
\qquad
\det\Phi_L=1.
\end{gathered}
\label{eq:t11-algebraic-cancellation-conditions}
\end{equation}

The same conditions that cancel $R_{\rm alg}$ also imply
\begin{equation}
\Phi_+=\Phi_-.
\end{equation}
The term containing the equation of motion in
\eqref{eq:t11-flatness-vs-eom} therefore vanishes identically:
\begin{equation}
\frac12(\Phi_--\Phi_+)E=0
\end{equation}
for arbitrary $E$.  Hence the resulting flatness condition does not imply
$E=0$.

\medskip

The reduced block-diagonal candidate
\eqref{eq:t11-validation-connection} is therefore not a genuine Lax
connection for the two-dimensional $T^{1,1}$ sigma model.  
Its flatness follows from an algebraic cancellation and does not encode
the equation of motion. In this sense, the distilled candidate is a
``fake'' Lax connection.
\section{Lax Pair for Geodesic Motion on \texorpdfstring{$T^{1,1}$}{T1,1}}
\label{sec:t11-geodesic}

So far, we have found that the block-diagonal connection fails to function as a genuine Lax representation of the full $T^{1,1}$ sigma model, which is non-integrable in the sense that chaotic motions emerge. However, it is worth considering the motion of a single particle whose target space is given by $T^{1,1}$\,. In this case, the base space becomes one-dimensional, and the resulting system is classically integrable.

\medskip 

Here, we shall discuss the Lax equation in this particle system by considering a particle-sector in the two-dimensional $T^{1,1}$ non-linear sigma model.

\subsection{Point-like Current Equation and Lax Pair}
\label{sec:point_like_reduction_of_Lax}

Starting from the current equation of motion derived in \eqref{eq:t11-current-eom}, we decompose the current in world-sheet coordinates $(\tau,\sigma)$ as
\begin{equation}J_\alpha=A_\alpha+K_\alpha,\qquad A_\alpha\in\mathfrak h,\qquad K_\alpha\in\mathfrak m,\qquad\alpha=\tau,\sigma .
\end{equation}
The equation of motion then takes the covariant form
\begin{equation}D_\tau K_\tau-D_\sigma K_\sigma=0\,,  
\label{eq:t11-worldsheet-current-eom}
\end{equation}
where the covariant derivative is defined as 
\begin{equation}
     D_\alpha X_{\beta} := \partial_\alpha X_\beta +[A_\alpha,X_{\beta}]
\end{equation}
for any $\mathfrak m$-valued world-sheet one-form component $X_\beta$\,.
The point-like sector, where all target-space coordinates depend solely on $\tau$, is characterized by the following conditions:
\begin{equation}\partial_\sigma X^i=0,\qquad J_\sigma=0,\qquad A_\sigma=K_\sigma=0.
\end{equation}
Consequently, the two-dimensional current equation \eqref{eq:t11-worldsheet-current-eom} reduces to
\begin{equation}
D_\tau K_\tau=0,
\label{eq:t11-reduced-eom}
\end{equation}
where $K_\tau=J_\tau^{(1)}$ and $A_\tau=J_\tau^{(0)}$.

\medskip 

The reduced equation of motion \eqref{eq:t11-reduced-eom} can be recast into a Lax pair formulation.\footnote{
Although the Liouville integrability of geodesic motion on
$T^{1,1}$ has been discussed in \cite{Babalic:2015integrability,Visinescu:2016t11},
here we present an explicit Lax-pair formulation in terms of the coset currents, given in \eqref{eq:t11-lax-pair}.
}
Consider the following ansatz for the Lax pair:
\begin{equation}
L=s\,K_\tau,\qquad M=A_\tau+c\,K_\tau,
\label{eq:t11-lax-pair}
\end{equation}
where $s$ and $c$ are constants. A direct calculation yields
\begin{align}
\frac{dL}{d\tau}-[L,M]&=s\frac{dK_\tau}{d\tau}-s\,[K_\tau,A_\tau]-sc\,[K_\tau,K_\tau]\nonumber\\ &=s\left(\frac{dK_\tau}{d\tau}+[A_\tau,K_\tau]\right)=s\,D_\tau K_\tau.
\label{eq:t11-lax-current-equivalence}
\end{align}
Since the term proportional to $c$ vanishes identically due to $[K_\tau,K_\tau]=0$, the Lax equation for $s\neq0$ is strictly equivalent to the equation of motion:
\begin{equation}\frac{dL}{d\tau}=[L,M]\qquad\Longleftrightarrow \qquad D_\tau K_\tau=0.\label{eq:t11-lax-equation}
\end{equation}
Note that the constant $c$ represents a flat direction of this one-dimensional Lax equation, rather than a true spectral parameter for the full two-dimensional $T^{1,1}$ sigma model.

\subsection{Coordinate Form of the Geodesic Equations}

Using the parametrization (\ref{para}) with the coordinates $(\theta_1, \phi_1, \theta_2, \phi_2, \psi)$ for the coset (\ref{eq:t11-coset}), the point-like currents are decomposed as
\begin{align}
K_\tau &= -\sin\theta_1\,\dot{\phi}_1\,K_1 +\dot{\theta}_1\,K_2 +\sin\theta_2\,\dot{\phi}_2\,L_1 +\dot{\theta}_2\,L_2 +\frac{2}{3}\eta\,T, \nonumber\\
A_\tau &= \frac12\left(\cos\theta_1\,\dot{\phi}_1 -\cos\theta_2\,\dot{\phi}_2\right)T_1 - \left[\frac16\left(\cos\theta_1\,\dot{\phi}_1 +\cos\theta_2\,\dot{\phi}_2\right) +\frac23\dot{\psi}\right]T_2 ,
\label{eq:t11-point-current-components}
\end{align}
where we have defined $\eta := \dot{\psi} + \cos\theta_1\,\dot{\phi}_1 + \cos\theta_2\,\dot{\phi}_2$, and ``~$\cdot$~'' denotes the $\tau$-derivative. 

The corresponding one-dimensional Lagrangian is given by
\begin{equation}
L_{1d} = \frac{1}{6}\bigl(\dot{\theta}_1^2 + \sin^2\theta_1\,\dot{\phi}_1^2 + \dot{\theta}_2^2 + \sin^2\theta_2\,\dot{\phi}_2^2\bigr) + \frac{1}{9}\eta^2.
\label{eq:t11-lagrangian}
\end{equation} 
Taking variations of this Lagrangian yields the coordinate equations of motion (EOMs):
\begin{align}
\ddot{\theta}_1 - \sin\theta_1\cos\theta_1\,\dot{\phi}_1^2 + \frac{2}{3}\sin\theta_1\,\dot{\phi}_1\,\eta &= 0, \label{eq:t11-eom-theta1}\\
\ddot{\theta}_2 - \sin\theta_2\cos\theta_2\,\dot{\phi}_2^2 + \frac{2}{3}\sin\theta_2\,\dot{\phi}_2\,\eta &= 0, \label{eq:t11-eom-theta2}\\
\frac{d}{d\tau}\bigl[\sin^2\theta_1\,\dot{\phi}_1 + \frac{2}{3}\eta\cos\theta_1\bigr] &= 0, \label{eq:t11-eom-phi1}\\
\frac{d}{d\tau}\bigl[\sin^2\theta_2\,\dot{\phi}_2 + \frac{2}{3}\eta\cos\theta_2\bigr] &= 0, \label{eq:t11-eom-phi2}\\
\dot{\eta} &= 0. \label{eq:t11-constraint}
\end{align}
The last equation dictates that $\eta$ is a conserved quantity along geodesics, which effectively reduces the system's dimension to four.

\medskip 

We now demonstrate that these coordinate EOMs are entirely encoded in the covariant current equation $D_\tau K_\tau = 0$. By evaluating the commutator $[A_\tau, K_\tau]$, we can expand $D_\tau K_\tau$ in the basis of the coset generators:
\begin{equation}
D_\tau K_\tau = E_{K_1}K_1 + E_{K_2}K_2 + E_{L_1}L_1 + E_{L_2}L_2 + E_T\,T.
\end{equation}
Direct calculation gives the explicit form of these coefficients:
\begin{align}
E_{K_1} &= -\frac{d}{d\tau}\!\left(\sin\theta_1\,\dot{\phi}_1\right) - \Omega_1\dot{\theta}_1, &
E_{K_2} &= \ddot{\theta}_1 - \Omega_1\sin\theta_1\,\dot{\phi}_1, \nonumber\\
E_{L_1} &= \frac{d}{d\tau}\!\left(\sin\theta_2\,\dot{\phi}_2\right) - \Omega_2\dot{\theta}_2, &
E_{L_2} &= \ddot{\theta}_2 + \Omega_2\sin\theta_2\,\dot{\phi}_2, \nonumber\\
E_T &= \frac{2}{3}\dot{\eta}, & &
\label{eq:t11-current-components}
\end{align}
with the effective angular velocities defined as 
\begin{equation}
    \Omega_1 := \cos\theta_1\,\dot{\phi}_1-\frac{2}{3}\eta, \quad \Omega_2 := -\cos\theta_2\,\dot{\phi}_2+\frac{2}{3}\eta.
\end{equation}
Substituting $\Omega_1$ and $\Omega_2$ into $E_{K_2}$ and $E_{L_2}$, we find
\begin{align}
E_{K_2} &= \ddot{\theta}_1 -\sin\theta_1\cos\theta_1\,\dot{\phi}_1^2 +\frac{2}{3}\sin\theta_1\,\dot{\phi}_1\eta, \label{eq:t11-current-theta1}\\
E_{L_2} &= \ddot{\theta}_2 -\sin\theta_2\cos\theta_2\,\dot{\phi}_2^2 +\frac{2}{3}\sin\theta_2\,\dot{\phi}_2\eta. \label{eq:t11-current-theta2}
\end{align}

The equivalence between the covariant equation and the coordinate EOMs is now manifest. The conditions $E_{K_2}=0$, $E_{L_2}=0$, and $E_T=0$ directly reproduce \eqref{eq:t11-eom-theta1}, \eqref{eq:t11-eom-theta2}, and the constraint \eqref{eq:t11-constraint}, respectively. Furthermore, the cyclic-coordinate equations \eqref{eq:t11-eom-phi1} and \eqref{eq:t11-eom-phi2} are obtained via the linear combinations:
\begin{align}
-\sin\theta_1\,E_{K_1}+\cos\theta_1\,E_T &= \frac{d}{d\tau}\left(\sin^2\theta_1\,\dot{\phi}_1 +\frac{2}{3}\eta\cos\theta_1\right), \label{eq:t11-phi1-from-current}\\
\sin\theta_2\,E_{L_1}+\cos\theta_2\,E_T &= \frac{d}{d\tau}\left(\sin^2\theta_2\,\dot{\phi}_2 +\frac{2}{3}\eta\cos\theta_2\right). \label{eq:t11-phi2-from-current}
\end{align}
Thus, away from the coordinate singularities ($\sin\theta_i=0$), $D_\tau K_\tau = 0$ is strictly equivalent to the full set of geodesic equations \eqref{eq:t11-eom-theta1}--\eqref{eq:t11-constraint}. Combined with \eqref{eq:t11-lax-equation}, this establishes that the one-dimensional Lax equation captures the coordinate geodesic motion on $T^{1,1}$.

\subsection{Numerical Check by Optimization}

It is interesting to verify the above analytical result by using the ML technique. 

\medskip 

As a numerical consistency check, we also train the structured ansatz
\begin{equation}
L=s\,K_\tau,
\qquad
M=\alpha\,A_\tau+\beta\,K_\tau
\label{eq:t11-geodesic-ml-ansatz}
\end{equation}
against the residual $\dot L-[L,M]$ using on-shell geodesic data generated from
\eqref{eq:t11-lagrangian}.  The analytic result \eqref{eq:t11-lax-pair}
predicts $\alpha=1$, while the overall scale $s$ is fixed by a mild
normalization penalty.  The parameter $\beta$ is invisible to this residual,
because $[K_\tau,K_\tau]=0$ for a single current.

\medskip

The training reproduces this structure: starting from
$(s,\alpha,\beta)=(0.35,0.25,0.4)$, it converges to
\begin{equation}
s=1.000,
\qquad
\alpha=1.000,
\qquad
\beta=0.400.
\end{equation}
This check confirms the implementation of the mechanical Lax equation.

\section{Summary and Discussion}
\label{sec:summary}

In this paper, we have investigated the use of machine learning to search for
Lax structures in two-dimensional non-linear sigma models.  The search is
performed at the level of local current data satisfying the equations of motion
and the flatness condition for currents, and the flatness of a candidate connection is
used as the optimization objective.

\medskip

We first considered the principal chiral model as the basic benchmark.  With a
minimal current-level ansatz, the trained coefficients do not converge to a
single isolated point, but instead lie on the expected spectral-parameter curve.
This shows that the spread of learned parameters should be interpreted as the
reconstruction of a Lax family rather than as a numerical ambiguity.  We also
introduced an explicit spectral-parameter input and trained coefficient
functions of the spectral parameter.  After fixing the reparametrization freedom
of the spectral coordinate, the network successfully reconstructs the expected
coefficient function over a region of the complex spectral-parameter plane.

\medskip

We then applied the same strategy to symmetric coset sigma models, focusing on
$S^2=SU(2)/U(1)$.  The learned coefficients reproduce the standard
symmetric-coset Lax family: the gauge part is fixed to its canonical
normalization, while the coset coefficients lie on the one-parameter locus
$bd=1$.  This provides a second benchmark, now involving the coset
decomposition, and confirms that the method can recover the spectral-parameter
structure of known integrable sigma models. 
In addition, we promoted the four coefficients to functions of an explicit
spectral parameter, without prescribing which of them carries it.  The runs
that reach the family return $a=c=1$ together with two mutually inverse
monomials, $b\propto\lambda^{n}$ and $d\propto\lambda^{-n}$ with $|n|=1$.
This structure is learned rather than imposed.  A more effective training
strategy and a more flexible ansatz are left to future work.

\medskip

For the non-symmetric coset $T^{1,1}$, the situation is different.  The direct
analogue of the symmetric-coset Lax ansatz collapses to the trivial current
connection and does not yield a nontrivial spectral-parameter family.  Enlarging
the search space with neural maps acting on the coset currents nevertheless
leads to reproducible low-loss configurations.  Inspection of the learned maps
reveals a simple block-diagonal structure, which can be distilled into a
finite-dimensional ansatz and retrained independently.  Thus, even in a case
where no suitable Lax ansatz is assumed in advance, the optimization provides a
compact analytic candidate worthy of further study.

\medskip

The analytic verification of this candidate is essential.  For the full
two-dimensional $T^{1,1}$ sigma model, the block-diagonal candidate cancels the
algebraic remainder, but the coefficient multiplying the
equation of motion vanishes at the same time.  Its flatness therefore does not
imply the sigma-model equations of motion, and the candidate should not be
regarded as a genuine Lax representation of the two-dimensional theory.  This
example shows explicitly that a small flatness loss, even when reproducible and
structurally simple, is not by itself evidence for the existence of a genuine
Lax connection.

\medskip

By contrast, the point-particle reduction on $T^{1,1}$ does admit a mechanical
Lax formulation.  The reduced current equation $D_\tau K_\tau=0$ is equivalent
to the geodesic equations away from coordinate singularities, and it is encoded
by a finite-dimensional Lax equation.  A structured optimization of the residual
$\dot L-[L,M]$ reproduces this mechanical Lax form, providing a numerical
consistency check of the analytic reduction.

\medskip

A natural extension of the present framework would be to combine Lax-based searches with conserved-charge-based diagnostics. Machine-learning approaches have demonstrated that conserved quantities can be identified from trajectory data or directly from differential equations ~\cite{Liu:2020aipoincare,Liu:2022aipoincare2}, while the Poisson-algebraic structure of conserved generators can also be learned~\cite{Hou:2024mlsd}. Such charge-based diagnostics may provide complementary information beyond the flatness residual when assessing a candidate Lax representation.

\medskip

Overall, the results demonstrate both the usefulness and the limitation of
machine-learning searches for Lax structures.  In controlled integrable
examples, the method can recover spectral-parameter families of Lax
connections.  In less understood cases, it can propose compact and
interpretable ans\"atze that guide analytic calculations.  However, the final
criterion remains analytic: low flatness loss alone does not certify the
existence of a genuine Lax connection or classical integrability.

\subsection*{Acknowledgments}

We would like to thank N.~Kubo for insightful discussions during the early stage.
Writing and running the numerical codes used in this paper, preparation of the figures, and drafting the manuscript were supported by ChatGPT, OpenAI Codex and Anthropic Claude Code.
The work of O.~F. was supported by RIKEN Special Postdoctoral Researchers Program.
The work of T.~S.~was supported by JST SPRING, Grant Number JPMJSP2110. 
The work of N.~T.~was supported in part by JSPS KAKENHI Grant No. JP22H05111 and 25K07282, and also by Kyoto University-Shimadzu Corporation Comprehensive Partnership.
The work of K.~Y.~was supported in part by JSPS KAKENHI Grant No.~JP22H05115, 25K07313 and the Asahipen Hikari Foundation.

\appendix

\section{Explicit spectral-parameter learning in the PCM: setup and diagnostics}
\label{app:pcm-lambda}

This appendix collects the details of the results in
section~\ref{sec:pcm-lambda-body} and shown in Fig.~\ref{fig:pcm-lambda-body}:
learning the coefficient functions $a(\lambda),c(\lambda)$ of the PCM
ansatz \eqref{eq:pcm-ac-ansatz} by minimizing the flatness residual alone, with
$\lambda$ supplied as an explicit network input. 
We fix the setup and the
objective, then define the diagnostics used to quantify the result.

\subsection{Setup and objective}
\label{app:pcm-setup}

The ansatz is \eqref{eq:pcm-ac-ansatz}, $L_+=a(\lambda)J_+$,
$L_-=c(\lambda)J_-$, with the coefficient network $\mathcal N_\theta$ of
Table~\ref{tab:pcm-net}.
As explained in section~\ref{sec:pcm-lambda-body}, flatness alone leaves a
reparametrization freedom because we have one complex equation for two complex unknowns at
each $\lambda$.
We fix this freedom by prescribing one coefficient analytically,
 $a(\lambda)= \lambda$, and learning only $c(\lambda)$.
If the training is successful, the learned $c(\lambda)$ should satisfy the
curve \eqref{eq:pcm-ac-curve}, equivalently
\begin{equation}
  c(\lambda) \to c_\text{an}(\lambda) =\frac{\lambda}{2\lambda-1}\,.
  \label{eq:pcm-can}
\end{equation}

\begin{table}[htbp]
\centering
\caption{The coefficient network $\mathcal N_\theta$. Only $c(\lambda)$ is
carried by the network; $a$ is prescribed analytically as $a(\lambda)=\lambda$,
as explained below.}
\label{tab:pcm-net}
\begin{tabular}{ll}
\hline
  Component & Specification \\
\hline
  Input & $(\operatorname{Re}\lambda,\operatorname{Im}\lambda)\in\mathbb{R}^2$ \\
  Hidden layers & 2 layers, 64 units each, $\tanh$ \\
  Output & $c(\lambda)\in\mathbb{C}$ (2 real outputs) \\
  Precision & float64 / complex128 \\
\hline
\end{tabular}
\end{table}

Training is carried out on the rectangle
\begin{equation}
  \mathcal R=\bigl\{\lambda\in\mathbb{C}\;:\;
  \operatorname{Re}\lambda\in[-2,0],\;\operatorname{Im}\lambda\in[-1,1]\bigr\}\,,
  \label{eq:pcm-domain}
\end{equation}
chosen so that the pole of the analytic solution \eqref{eq:pcm-can} at
$\lambda=1/2$ lies outside it.
Each optimization step draws $64$ values of $\lambda$
uniformly from $\mathcal R$ together with a batch of $1024$ on-shell current
configurations, and averages the objective over both. The analytic solution
\eqref{eq:pcm-can} is never used in training; it enters only as the evaluation
diagnostic \eqref{eq:pcm-cabs} below.

\medskip

The objective is the component-norm normalized flatness
\eqref{eq:pcm-loss-cn} of section~\ref{sec:pcm-lambda-body}, here with
$\varepsilon=10^{-10}$. Its role is 
to equalize the training signal over current configurations whose amplitudes
vary widely across the batch.\footnote{Here the trivial solution $a=c=0$ is
already excluded by the prescription $a(\lambda)=\lambda$. 
If we do not impose such
a gauge fixing, the normalization also serves as a disincentive against collapse
onto the trivial solution: the numerator and the denominator scale together
under $L_\pm\to s\,L_\pm$, so shrinking the connection to zero amplitude is not
rewarded. This mechanism becomes active in the $S^2$ search of
section~\ref{sec:priorfree-body}.}
As we show at the end of this appendix, in the
present setup the same normalization equalizes the signal over the spectral
parameter as well.

\medskip

The data term actually minimized is \eqref{eq:pcm-loss-cn}
as it stands, formed
sample by sample from the currents, and not its on-shell reduction below. 
It is nevertheless instructive to evaluate
it on the training data: every term of $F_{+-}$ and of the denominator is then
proportional to $[J_+,J_-]$ by \eqref{eq:pcm-eom-mc-combined}, the common factor
$\|[J_+,J_-]\|^2$ cancels between the two, and
\begin{equation}
  \mathcal L_{\rm cn}\Bigl|_{\text{on shell}}
  \;\simeq\;
  \Bigl\langle \tfrac14\,\bigl|a+c-2ac\bigr|^2\big/D \Bigr\rangle_{\lambda}\,,
  \qquad
  D=\tfrac14|a|^2+\tfrac14|c|^2+|ac|^2\,,
  \label{eq:pcm-cn-onshell}
\end{equation}
where $\simeq$ is an equality up to the regulator $\varepsilon$ in the
denominator of \eqref{eq:pcm-loss-cn}. The right-hand side is a function of
$(a,c)$ alone; the average over currents has become trivial, leaving only the average over $\lambda$, so every sampled current configuration contributes to the gradient with the same strength, irrespective of its amplitude.

\medskip

Optimization uses Adam with a learning rate decreased smoothly from
$2\times10^{-3}$ to $2\times10^{-5}$ along a cosine profile, after a linear
ramp-up over the first $150$ steps, for $10^4$ steps, with gradients clipped to
norm $10$. No regularizer is added: the objective is the normalized flatness
residual \eqref{eq:pcm-loss-cn} alone.

\subsection{Diagnostics and result}
\label{app:pcm-results}

All diagnostics are evaluated after training on a $200\times200$ grid
$\{\lambda_k\}_{k=1}^{N}$ ($N=200^2$) covering the training rectangle
$\mathcal R$ of \eqref{eq:pcm-domain}, not merely on the real-axis slice
displayed in Fig.~\ref{fig:pcm-lambda-body}. From the
learned coefficients we form the spectral-curve gap
\begin{equation}
  g(\lambda)=\bigl|a(\lambda)+c(\lambda)-2\,a(\lambda)c(\lambda)\bigr|\,,
  \label{eq:pcm-gap}
\end{equation}
which measures the pointwise violation of \eqref{eq:pcm-ac-curve} (it depends on
$\lambda$ only, the on-shell current factor having cancelled), and report its
mean and maximum over the grid,
\begin{equation}
  \mathrm{gap\_mean}=\frac{1}{N}\sum_{k=1}^{N} g(\lambda_k)\,,\qquad
  \mathrm{gap\_max}=\max_{k} g(\lambda_k)\,.
  \label{eq:pcm-gap-stats}
\end{equation}
Since the parametrization is fixed by $a=\lambda$, the learned $c$ can in
addition be compared directly with the analytic solution \eqref{eq:pcm-can},
through the deviation
\begin{equation}
  \delta c(\lambda)=c(\lambda)-c_{\rm an}(\lambda)\,,
  \qquad
  \mathrm{abs\_mean}=\frac{1}{N}\sum_{k=1}^{N}\bigl|\delta c(\lambda_k)\bigr|\,,
  \qquad
  \mathrm{abs\_max}=\max_{k}\bigl|\delta c(\lambda_k)\bigr|\,,
  \label{eq:pcm-cabs}
\end{equation}
the genuine numerical-solution error, summarized by the same two statistics as
the gap.

\medskip

\begin{figure}[htbp]
  \centering
  \includegraphics[width=\textwidth]{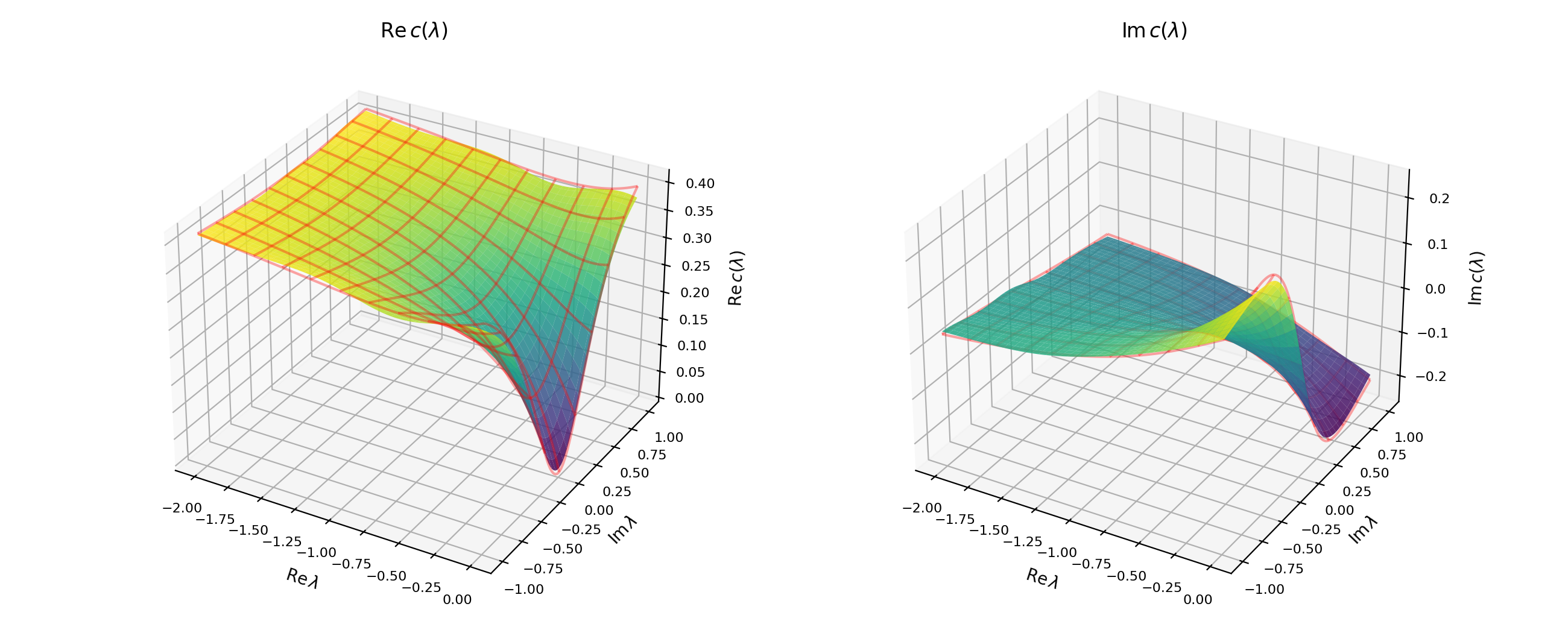}
  \caption{The learned coefficient $c(\lambda)$
    ($\operatorname{Re}c$, left; $\operatorname{Im}c$, right) of the run of
    Fig.~\ref{fig:pcm-lambda-body}, over an $80\times80$ grid on the training
    rectangle $\mathcal R$ of \eqref{eq:pcm-domain} in the complex
    $\lambda$-plane; the red wireframe is the analytic solution
    $c_{\rm an}=\lambda/(2\lambda-1)$ of \eqref{eq:pcm-can}, shown for
    comparison only. The partner
    coefficient $a$ is omitted because it is prescribed rather than learned,
    $a(\lambda)=\lambda$.}
  \label{fig:pcm-lambda-3d}
\end{figure}

The run shown in Fig.~\ref{fig:pcm-lambda-body} attains
\begin{align*}
  &\mathrm{gap\_mean}=9.5\times10^{-3},\qquad
  \mathrm{gap\_max}=5.3\times10^{-2},\\
  &\mathrm{abs\_mean}=3.0\times10^{-3},\qquad
  \mathrm{abs\_max}=2.4\times10^{-2}
\end{align*}
over $\mathcal R$. Figure~\ref{fig:pcm-lambda-3d} displays the learned
$c(\lambda)$ over that rectangle. Only $c$ is shown: $a$ is not an output of the
network but the prescribed function $a(\lambda)=\lambda$.
The learned $c(\lambda)$ tracks $\lambda/(2\lambda-1)$ over the whole
rectangle, confirming that the real-axis slice of
Fig.~\ref{fig:pcm-lambda-body} is not a projection artifact but a section
through a genuine function of the complex spectral parameter.

\subsection{Loss function structure and error distribution}

The on-shell form \eqref{eq:pcm-cn-onshell} of the loss function can be evaluated in closed form near
the solution, and doing so shows how the normalization acts in the spectral
parameter.
With $a$ pinned to
$a=\lambda$, the only learned quantity is $c$; writing $c=c_{\rm an}+\delta c$
with $c_{\rm an}$ of \eqref{eq:pcm-can}, the numerator factorizes exactly,
\begin{equation}
  a+c-2ac=\lambda+(1-2\lambda)\,c=(1-2\lambda)\,\delta c\,.
  \label{eq:pcm-gap-exact}
\end{equation}
The denominator, evaluated on the solution, is
\begin{equation}
  D=\frac{|\lambda|^2}{4}\,
    \frac{|2\lambda-1|^2+1+4|\lambda|^2}{|2\lambda-1|^2}\,.
  \label{eq:pcm-D-exact}
\end{equation}
which does degenerate, $D\approx\tfrac12|\lambda|^2\to 0$ as $\lambda\to0$.
Taken
alone this would make the effective per-$\lambda$ gain blow up at the origin.
But $c_{\rm an}$ vanishes at exactly the same rate,
$|c_{\rm an}|^2=|\lambda|^2/|2\lambda-1|^2$, and the two powers of $|\lambda|$
cancel identically once the deviation is measured relative to $c_{\rm an}$:
\begin{equation}
  \mathcal L_{\rm cn}\Bigl|_{\text{on shell}}
  \;=\;
  \Bigl\langle\,\kappa(\lambda)\,
    \bigl|\delta c(\lambda)/c_{\rm an}(\lambda)\bigr|^2
  \Bigr\rangle_{\lambda}\,,
  \qquad
  \kappa(\lambda)=\frac{|2\lambda-1|^2}{|2\lambda-1|^2+1+4|\lambda|^2}\,,
  \label{eq:pcm-cn-relative}
\end{equation}
up to corrections of higher order in $\delta c$ (and the regulator
$\varepsilon$). The weight $\kappa$ is mostly flat on $\mathcal R$: 
$\kappa\in[\tfrac12,\tfrac23]$ over the entire training rectangle,
where the minimum $1/2$ is attained along the edge $\operatorname{Re}\lambda=0$
(on which $|2\lambda-1|^2=1+4|\lambda|^2$ exactly) and the maximum $2/3$ at
$\lambda=-1/2$.

\medskip

What is actually minimized is therefore the \emph{relative} error of $c$,
weighted almost uniformly over the spectral-parameter plane. The component-norm
normalization thus equalizes the training signal in both directions at once:
across current configurations, by construction, and across $\lambda$ as well in this setup, as a
consequence of the fact that numerator and denominator degenerate together
where the solution does.
Neither the flatness of $\kappa$ nor the
cancellation behind it is automatic; both reflect the particular pair
$(a=\lambda,\,c_{\rm an})$.

\medskip

While the loss function has the structure explained above, the actual relative error distribution is not uniform over $\mathcal R$.
The modulus of $\delta c$ is
close to constant over $\mathcal R$: its mean in radial bins of $|\lambda|$
stays between $2.7\times10^{-3}$ and $3.8\times10^{-3}$. It implies that the relative
deviation grows toward the origin, from $7\times10^{-3}$ at $|\lambda|\simeq1$ to
$2.5\times10^{-2}$ at $|\lambda|\simeq0.15$ and beyond.
An even training signal
is thus necessary but not sufficient for an even relative accuracy.
The analytic solution~\eqref{eq:pcm-can} and Figure~\ref{fig:pcm-lambda-3d} show that the learned $c(\lambda)$ has a relatively large gradient around $\lambda\sim 0$, which would be a cause of the larger relative error there.
This problem may be fixed by using more sophisticated methods (e.g., adaptive sampling and a different network architecture) to better capture the behavior of $c(\lambda)$ near $\lambda=0$.
We defer such a technical investigation to future works.

\section{Explicit spectral-parameter learning for \texorpdfstring{$S^2$}{S2}: setup and diagnostics}
\label{app:priorfree-lambda}

This appendix collects the details of the construction presented in
section~\ref{sec:priorfree-body}: learning the coefficient functions $a\,(\lambda),b\,(\lambda),c\,(\lambda),d\,(\lambda)$ of the
symmetric-coset ansatz \eqref{eq:sym-abcd-ansatz} on a global annulus in the spectral parameter $\lambda$,
by minimizing the flatness residual together with a coefficient-agnostic term that forbids the
$\lambda$-independent charts. We fix the setup and the objective, including the control run that
delimits what the construction establishes, and then define the diagnostics and report the
per-seed results.

\subsection{Setup and objective}
\label{app:s2-setup}

The ansatz is the Laurent form \eqref{eq:s2-laurent} with $K=2$: each of $a,b,c,d$ is an
independent complex Laurent polynomial of degree $[-2,2]$, so the model carries $4\times5$
complex coefficients as its only parameters. There is no neural network here; unlike the PCM
case of Appendix~\ref{app:pcm-lambda} the $\lambda$-dependence is represented by an explicit
global functional form. Two considerations lead to this choice. First, we do not want to
prescribe which coefficient carries the spectral parameter $\lambda$. All four must therefore be
represented in the same way, and each must be free either to stay constant or to vary, so that
the choice is left to the flatness residual. 
Second, the coefficients of a Lax connection are generally meromorphic
functions of $\lambda$: both the standard $S^2$ family
\eqref{eq:sym-standard-lax}
and the PCM coefficients \eqref{eq:pcm-ac-lambda} are of this type.
For the standard $S^2$ family,
the relevant dependence is captured by Laurent monomials, so a truncated
Laurent expansion is the simplest finite-dimensional ansatz adapted to this
example.
Each coefficient is holomorphic on $\mathbb C^\times$ by construction,
while the negative-power
modes represent possible poles at $\lambda=0$ explicitly.
On the annular
training domain, monomials such as $\lambda^n$ are therefore represented
exactly, so maps with nonzero winding around the excluded point are already
contained in the ansatz. 
By contrast, a generic network taking
$(\operatorname{Re}\lambda,\operatorname{Im}\lambda)$ as input does not
impose holomorphicity or an explicit pole structure. The Laurent form also
makes the learned dependence directly interpretable mode by mode.

\medskip

The spectral parameter is drawn from the origin-centred annulus
\begin{equation}
  \mathcal{A}=\{\lambda\in\mathbb{C}\;:\;0.25\le|\lambda|\le4\},
  \label{eq:s2-domain}
\end{equation}
uniformly in $\log|\lambda|$ and in $\arg\lambda$. This measure is invariant under
$\lambda\to1/\lambda$, so the two branches $b\propto\lambda^{\pm1}$ are sampled on an equal
footing. Each step uses $256$ values of $\lambda$ against a batch of $8192$ on-shell current
configurations generated from \eqref{rel1}--\eqref{rel2}; training runs for $1.2\times10^{4}$
steps of Adam,
in float64/complex128,
with the learning rate decreased from $10^{-2}$ to
$10^{-5}$ along a cosine profile after a short linear ramp.

\medskip

The cutoff $K$ deserves a comment. Expressivity argues for taking it large, so that the
ansatz restricts as little as possible which functions of $\lambda$ the search can reach.
What prevents it is the conditioning of the Laurent basis on the annulus
\eqref{eq:s2-domain}, where the modulus of $\lambda^{k}$ ranges over
$[4^{-|k|},4^{|k|}]$. The largest and the smallest value taken by the basis functions of
the ansatz then differ by $4^{2K}$: a factor of $2.6\times10^{2}$ at $K=2$, and of
$4\times10^{9}$ at $K=8$. At high order the residual and its gradient are dominated by the
outermost modes at random initialization, and training diverges. This leaves a cutoff of a
few, and $K=2$ sits at the low end of that range. It is a limitation of this
parametrization rather than a considered choice, and lifting it requires a
better-conditioned representation of the $\lambda$-dependence.

\medskip

Two things should be kept apart in judging what the cutoff costs. In retrospect it costs
nothing here: if $b$ and $d=1/b$ are both finite Laurent series then $b$ is a pure power,
so the family \eqref{eq:sym-standard-lax} is contained exactly in the ansatz for every
$K\ge1$. That argument uses the answer, which a search that is agnostic about the solution
does not have. What such a search does have is the weight carried by the outermost modes.
A fit truncated by the cutoff pushes amplitude onto $|k|=K$, and amplitude found there is
the signal to raise $K$. In these runs it is not found there: the $|k|=2$ moduli of the
converged runs are of order $10^{-2}$ or smaller against dominant modes of order unity
(Table~\ref{tab:annulus-modes-full}). The modes are available rather than frozen, and they
carry $1.6\times10^{-2}$ in the least converged run, where they hold the residual of an
incomplete fit.

\medskip

The loss function for the training is \eqref{eq:nc-penalty}. Its first term is the component-norm residual
$\mathcal{L}_{\rm cn}$ of \eqref{eq:pcm-loss-cn}, evaluated here with the coset currents; being
scale-invariant it removes the incentive to shrink onto $L=0$,
exactly as in the PCM. The second is the non-constancy penalty $w_{\rm nc}\prod_f r_f$ of
\eqref{eq:nc-fraction} with $w_{\rm nc}=0.3$. The third is a low-order penalty of weight
$w_{\rm reg}=10^{-4}$ on the $|k|\ge2$ modes, a preference for the lowest available order.

\medskip

Alongside the search we run a \emph{control} with that second term switched off,
$w_{\rm nc}=0$, and everything else (annulus, ansatz, batch and $w_{\rm reg}$) left
unchanged. It is what shows that flatness and the low-order preference alone do not select a
non-constant chart. Its collapse is not a matter of insufficient training: as noted in
section~\ref{sec:priorfree-body}, a constant chart is an exact point of
\eqref{eq:sym-expected} and hence an exact minimum of the flatness residual, so we stop the
control at $4\times10^{3}$ steps against the $1.2\times10^{4}$ of the search.

\medskip

It is worth being explicit about what each of the last two does \emph{not} do:
neither distinguishes $k=+1$ from $k=-1$, so neither decides the sign of the winding; neither is
sensitive to the overall scale of a coefficient, so neither fixes the amplitude; and the
non-constancy term treats $a,b,c,d$ symmetrically, so it does not designate the coefficient that
must carry the spectral parameter. 
The symmetry between the two winding signs is exact.
On \eqref{eq:s2-laurent} the map $\lambda\to1/\lambda$ reverses every Laurent
mode, $f_k\mapsto f_{-k}$, and it leaves all three terms of
\eqref{eq:nc-penalty} invariant: the low-order penalty selects $|k|\le2$, the non-constancy fraction
reads the $k=0$ mode alone, and \eqref{eq:s2-domain} is sampled from an inversion-invariant measure.
On a grid closed under $\lambda\to1/\lambda$ the reflected and unreflected losses agree to
$4\times10^{-15}$, i.e.\ to roundoff. This is what makes the winding sign gauge. The $w_{\rm reg}$ term does prefer low
order; the regulator in \eqref{eq:nc-fraction} is $\varepsilon=10^{-24}$.

\subsection{Diagnostics and results}
\label{app:s2-results}

We record the following as the diagnostics of the training.
All diagnostics are evaluated after training on a polar grid of $48\times180$ points covering the annulus
$\mathcal{A}$, using a batch of on-shell currents independent of the training batch.

\begin{itemize}

\item \textbf{Laurent mode moduli.} The moduli $|f_k|$ of the learned modes, which show
how the chart represents the one-parameter family \eqref{eq:sym-expected}, and the dominant mode
$k^{\ast}_f=\arg\max_{k}|f_k|$ of each coefficient.

\item \textbf{Non-constancy fractions.} The value of $r_f$ of \eqref{eq:nc-fraction} after
training, which identifies the coefficients that left the constant mode. It is tabulated for
$b$ and $d$ in Table~\ref{tab:annulus-modes-full}; for $a$ and $c$ it stays at $1.000$ in every
run. It is the one quantity here that also enters the loss.

\item \textbf{Deviation from the one-parameter family.} From the learned coefficients we form
\begin{equation}
  \bigl\langle|bd-1|\bigr\rangle,\qquad \max|a-1|,\qquad \max|c-1|,
  \label{eq:s2-familyerr}
\end{equation}
the deviation from \eqref{eq:sym-expected}, which measures whether the chart lies on
the solution set. The family is used here only as an evaluation diagnostic, never as a target:
it appears nowhere in the loss \eqref{eq:nc-penalty}.

\item \textbf{Contamination.} How far each coefficient is from a pure monomial,
\begin{equation}
  \rho_f
  =
  1-\frac{|f_{k_f^{\ast}}|^{2}}{\sum_{k}|f_k|^{2}}
  \in[0,1],
  \label{eq:contamination}
\end{equation}
the fraction of the squared coefficient norm carried by modes other
than the dominant one. Equivalently, $1-\rho_f$ is the analogue of
\eqref{eq:nc-fraction} with the dominant mode $k_f^\ast$ in place of
the constant mode and with the regulator omitted. Thus $\rho_f$
vanishes precisely on a monomial of any degree.
Unlike $r_f$ it is a diagnostic only and never enters the
loss. Being blind to the degree, $\rho_f$ is meaningful only
alongside $k^{\ast}_f$, and it compares runs only among those sharing a dominant mode: a
constant is a monomial as well, so an exactly collapsed chart would have $\rho_f=0$. Telling
a collapsed chart from a non-constant one is the task of $r_f$, not of $\rho_f$.
For the summary in Table~\ref{tab:annulus-modes}, we report the larger
contamination of the two coefficients carrying the spectral dependence,
\begin{equation}
\rho:=\max\{\rho_b,\rho_d\}.
\label{eq:summary-contamination}
\end{equation}

\end{itemize}

\begin{table}[htbp]
\centering
\caption{Laurent mode moduli $|f_k|$ of $f=b$ and $f=d$ for the control and the
  three runs of Table~\ref{tab:annulus-modes}, the full version of
  Table~\ref{tab:priorfree-learned}.
  The dominant mode of each coefficient is set in bold; its amplitude is gauge, and
  only the product of the two is fixed, here to unity with an error of $5.1\times10^{-7}$. The last column
  is the non-constancy fraction $r$ of \eqref{eq:nc-fraction}, the share of each row
  carried by its $k=0$ entry, and the term the loss penalizes.
  $a$ and $c$ sit on their $k=0$ mode with modulus $1.000$ throughout and are
  omitted. 
  }
\label{tab:annulus-modes-full}
\begin{tabular}{llccccc|c}
\hline
run & & $k{=}{-}2$ & $k{=}{-}1$ & $k{=}0$ & $k{=}{+}1$ & $k{=}{+}2$ & $r$ \\
\hline
control & $b$ & $3.8\times10^{-3}$ & $7.4\times10^{-2}$ & $\mathbf{0.7322}$ & $7.2\times10^{-2}$ & $3.6\times10^{-3}$ & $0.980$ \\
 & $d$ & $6.9\times10^{-3}$ & $1.4\times10^{-1}$ & $\mathbf{1.3422}$ & $1.3\times10^{-1}$ & $6.5\times10^{-3}$ & $0.980$ \\
\hline
1 & $b$ & $3.5\times10^{-4}$ & $\mathbf{1.0057}$ & $5.3\times10^{-4}$ & $2.8\times10^{-7}$ & $<\!10^{-9}$ & $2.7\times10^{-7}$ \\
 & $d$ & $<\!10^{-9}$ & $1.2\times10^{-7}$ & $3.5\times10^{-4}$ & $\mathbf{0.9943}$ & $5.2\times10^{-4}$ & $1.2\times10^{-7}$ \\
\hline
2 & $b$ & $1.2\times10^{-2}$ & $\mathbf{0.4916}$ & $1.7\times10^{-7}$ & $<\!10^{-9}$ & $<\!10^{-9}$ & $1.3\times10^{-13}$ \\
 & $d$ & $2.9\times10^{-5}$ & $1.2\times10^{-3}$ & $4.9\times10^{-2}$ & $\mathbf{2.0341}$ & $7.4\times10^{-7}$ & $5.9\times10^{-4}$ \\
\hline
3 & $b$ & $1.6\times10^{-2}$ & $\mathbf{0.4375}$ & $3.2\times10^{-7}$ & $1.6\times10^{-8}$ & $8.4\times10^{-9}$ & $5.2\times10^{-13}$ \\
 & $d$ & $1.0\times10^{-4}$ & $2.9\times10^{-3}$ & $8.2\times10^{-2}$ & $\mathbf{2.2859}$ & $2.0\times10^{-6}$ & $1.3\times10^{-3}$ \\
\hline
\end{tabular}
\end{table}

\begin{table}[htbp]
\centering
\caption{Results on the annulus \eqref{eq:s2-domain}. The control of
  Appendix~\ref{app:s2-setup} ($w_{\rm nc}=0$, $4\times10^{3}$ steps) is on the first
  line, and collapses for each of the three seeds; runs $1$--$3$ differ only in the random
  initialization, are ordered by
  $\mathcal{L}_{\rm cn}$, and are trained for $1.2\times10^{4}$ steps.
  The three seeds were fixed in advance and all four runs share one batch of on-shell
  currents; nothing was selected after the fact.
  In each of them the coefficients that leave the constant mode
  are $b$ and $d$ and only these ($r_a=r_c=1.000$ throughout), although the penalty names
  none of them; the resulting $k^{\ast}$ are gauge-equivalent to the standard assignment
  by $\lambda\to1/\lambda$. That all three land on the same sign is a coincidence of these
  seeds: on a larger set of initializations the two signs occur equally often, six
  against five among the charts of the sweep reported below. Read $\rho$ together with $k^{\ast}$: the low contamination of
  the control and of the three runs means convergence onto a constant and onto the
  $k=\pm1$ modes respectively (Table~\ref{tab:annulus-modes-full}). The exact values are
  $\mathcal{L}_{\rm cn}=0$, $|bd-1|=0$ and $\rho=0$. Run $1$ is the chart of
  Table~\ref{tab:priorfree-learned}.}
\label{tab:annulus-modes}
\begin{tabular}{lcccccc}
\hline
run & $\mathcal{L}_{\rm cn}$ & $\langle|bd-1|\rangle$ & $\max|a-1|$ & $k^{\ast}_b$ & $k^{\ast}_d$ & $\rho$ \\
\hline
control & $4.3\times10^{-7}$ & $1.3\times10^{-3}$ & $9.1\times10^{-5}$ & $0$ & $0$ & $2.0\times10^{-2}$ \\
\hline
1 & $3.3\times10^{-15}$ & $2.1\times10^{-7}$ & $8.9\times10^{-8}$ & $-1$ & $+1$ & $4.0\times10^{-7}$ \\
2 & $4.3\times10^{-11}$ & $9.6\times10^{-6}$ & $2.5\times10^{-6}$ & $-1$ & $+1$ & $5.9\times10^{-4}$ \\
3 & $9.9\times10^{-10}$ & $4.3\times10^{-5}$ & $6.3\times10^{-6}$ & $-1$ & $+1$ & $1.3\times10^{-3}$ \\
\hline
\end{tabular}
\end{table}

Tables~\ref{tab:annulus-modes-full} and~\ref{tab:annulus-modes} collect the outcome. The
qualitative result is common to the three runs: all escape the constant chart, do so
through $b$ and $d$ alone, keep $a$ and $c$ pinned to unity, and land on the single mode $k=-1$
of $b$ against $k=+1$ of $d$. The quantitative accuracy is not: the three span six orders
of magnitude in $\mathcal{L}_{\rm cn}$, and only the best approaches the exact monomial
closely. The residual in the weaker ones is a small admixture of the $k=-2$ mode into $b$, for run
$3$ a modulus of $1.6\times10^{-2}$ against a dominant mode of $0.44$. 
The summary contamination
\eqref{eq:summary-contamination} measures this admixture, falling from
$\rho=1.3\times10^{-3}$ in run $3$ to $4.0\times10^{-7}$ in run $1$;
it orders the runs in the same way as $\mathcal{L}_{\rm cn}$.
The convergence is slow rather than stalled: all three runs were
still improving when training stopped, having gained between one and five orders
of magnitude on going from $4\times10^{3}$ to $1.2\times10^{4}$ steps. The gain is a trend, not a
decrease at every step: each step draws a fresh $\lambda$ minibatch, so the residual fluctuates by a
factor of a few along the way.

\medskip

All three runs of Table~\ref{tab:annulus-modes} end on a chart. That is not the typical
outcome. A sweep of $20$ runs from fresh
initializations, stopped at $2\times10^{3}$ steps, gave $11$ charts with
$\mathcal{L}_{\rm cn}<10^{-4}$ and $9$ runs stalled near $10^{-2}$. In a stalled run only one of the
four coefficients becomes $\lambda$-dependent, whereas a chart needs $b$ and $d$ to move together
since $bd=1$. Training longer does not help: three stalled runs continued to the
$1.2\times10^{4}$ steps of Table~\ref{tab:annulus-modes} stayed at
$\mathcal{L}_{\rm cn}\sim10^{-2}$, with $\langle|bd-1|\rangle$ between $0.09$ and $1.0$. The
outcome is set by the initialization, not by the on-shell currents a run is trained on: repeating a
run against independent batches of currents reproduces its residual to ten percent and never turns
a stalled run into a chart.

\medskip

This failure is due to the form of the non-constancy penalty in \eqref{eq:nc-penalty}. The penalty is
the product $\prod_f r_f$ over the four coefficients, so it asks for one non-constant coefficient and
no more. Suppose for example that $a$ is the coefficient that becomes $\lambda$-dependent first. The
penalty is then satisfied by $a$ alone and asks for nothing further, and what is left to complete the
chart is the flatness residual.
The flatness
residual pulls $a$ back towards a constant ($a=1$), while the penalty pushes the other way, since it switches on again as $a$
tends to a constant.
The run settles where the two balance, at which the residual becomes order
$10^{-2}$ and stays there however long the training continues.
Removing failures of this kind needs a better loss function and training strategy,
which we leave to future work.


\providecommand{\href}[2]{#2}\begingroup\raggedright\endgroup

\end{document}